\documentclass[useAMS,usenatbib]{mn2e}
\usepackage{graphicx}

\def\la{\raise.5ex\hbox{$<$}\kern-.8em\lower 1mm\hbox{$\sim$}}
\def\ma{\raise.5ex\hbox{$>$}\kern-.8em\lower 1mm\hbox{$\sim$}}

\def\kms{$\rm km\, s^{-1}$}
\def\cm3{$\rm cm^{-3}$}
\def\Ts{$\rm T_{*}$~}
\def\Vs{$\rm V_{s}$~}
\def\n0{$\rm n_{0}$}
\def\B0{$\rm B_{0}$}

\def\Ne{$\rm N_{e}$}
\def\Te{$\rm T_{e}$}

\def\erg{$\rm erg\, cm^{-2}\, s^{-1}$}
\def\mum{$\mu$m~}
\def\muu{$\mu$}

\def\L12{L$_{12\mu m}$~}
\def\12{L$_{12\mu m}$~}
\def\F12{F$_{12\mu m}$~}
\def\agr{a$_{gr}$}
\def\Hb{H${\beta}$~}
\def\Ha{H${\alpha}$~}
\def\Hg{H${\gamma}$~}

\def\RO3{R$_{[OIII]}$}

\title[Modelling SGRB100628A host spectra]{AGN and starburst coexistence in the short GRB100628A 
host galaxy.
}

\author[M. Contini]{M. Contini 
\\
School of Physics and Astronomy, Tel Aviv University, Tel Aviv
69978, Israel \\
}

\begin{document}


\pagerange{\pageref{firstpage}--\pageref{lastpage}} \pubyear{2009}

\maketitle

\label{firstpage}

\begin{abstract}
We have modelled the  line ratios  and the continuum spectral energy distribution (SED)
of  the short gamma-ray burst GRB100628A host galaxy complex.
The results suggest that an active galactic nucleus (AGN)  and a  star-burst (SB) coexist in 
Nicuesa Guelbenzu et al.'s 'galaxy C'.
The AGN spectrum  is explained by a relatively strong flux.
Two of the observed  regions (blobs)  are located in the ISM, outside the  AGN photoionization cones.
 The  strongest \Hb line flux  calculated by a SB dominated model  was found  in the tidal
tail of galaxy C, while the strongest \Hb line  flux calculated by an AGN dominated model  was found
in the galaxy C bulk.  O/H element abundances are near solar everywhere, while N/H  spans  a factor of $\sim$10.
The  radio data have different origins, thermal bremsstrahlung and reprocessed radiation by dust.
The infrared-optical photometric data measured from  Nicuesa Guelbenzu et al.'s 'galaxy D', located
 outside the X-ray telescope (XRT) error circle, are reproduced by
a black-body  flux  at 3000 K.  This  flux   may   represent the radiation of the  underlying old star population 
observed throughout  the SED in the infrared.  AGNs  in other SGRB hosts  are investigated.

\end{abstract}


\section{Introduction}

 The observations show that short duration GRB (SGRB) last less than 2s, while long GRB (LGRB) have longer periods
(Kouvelioutou et al 1993).   New evidence in support of the compact binary merger hypothesis
was provided in particular by the observation of a kilonova/macronova accompanying SGRB130603B, and subsequently
the observed gamma-ray flash accompanying GW170817
 (e.g. Levan et al 2017,  Villar et al 2017).
 At present, some other characteristics of SGRB are not definitive.
For instance,  short GRB host galaxy  types -
 on the basis of the finding  of an elliptical host galaxy
(Berger et al. 2005)- 
indicate that  some  SGRBs arise from an old stellar population
(Bloom et al. 2005, Gehrels et al. 2005) consistent with the hypothesis  that compact object binaries  are their progenitors.
Fong \& Berger (2013)  by  HST observations and a detailed analysis of 22 short GRB host galaxies 
 combined with the results from Fong et al. (2010)
concluded   that $\sim$ 1/4 of SGRB host galaxies  are elliptical
with  median effective sizes  of  $\sim$3.6 kpc
 and  that 30  - 45  percent 
are located on the faintest optical regions of their host galaxies, while $\sim$ 55 percent  occur in the faintest UV regions.
Fong \& Berger (2010)  claim that short GRB do not occur 
in regions of star formation or stellar mass
but  the SGRB progenitor systems  should "migrate from their birth site to the
 eventual explosion sites, implying kicks in compact binary systems (NS (neutron star)-NS, NS-BH (black hole))
with velocities of 20-140 \kms".
"Kicks" were invoked already in the early '70ies  to explain supernova (SN) remnants hosting an associated pulsar displaced 
 from the centre or even missing (e.g. Vela SNR, van den Bergh et al 1973).
Progenitor migration throughout the SGRB system  host galaxies led Nicuesa Guelbenzu et al (2015) to search 
for the SGRB100628A host
within  a mixed complex of galaxies spiral and elliptical at different redshifts,  stars, 
AGN,  starbursts , etc.  which are all close to the GRB location based on its X-ray detection.

\begin{figure*}
\centering
\includegraphics[width=16.2cm]{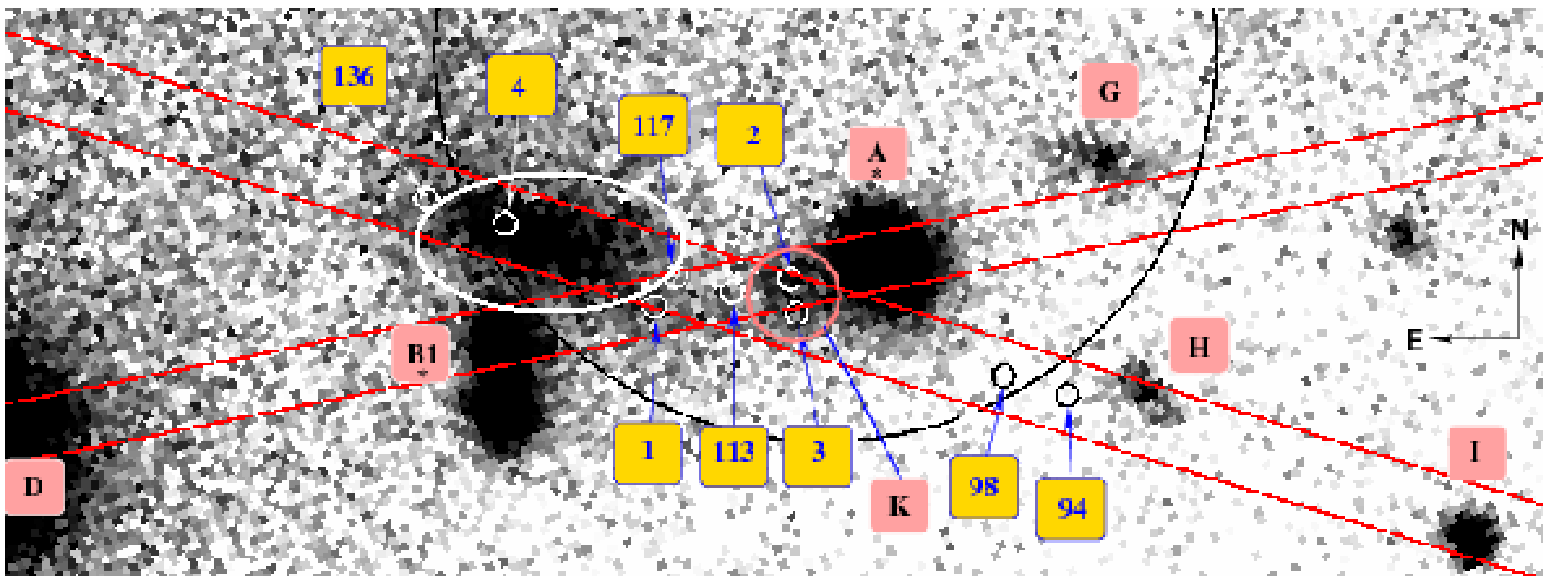}
\includegraphics[width=16.2cm]{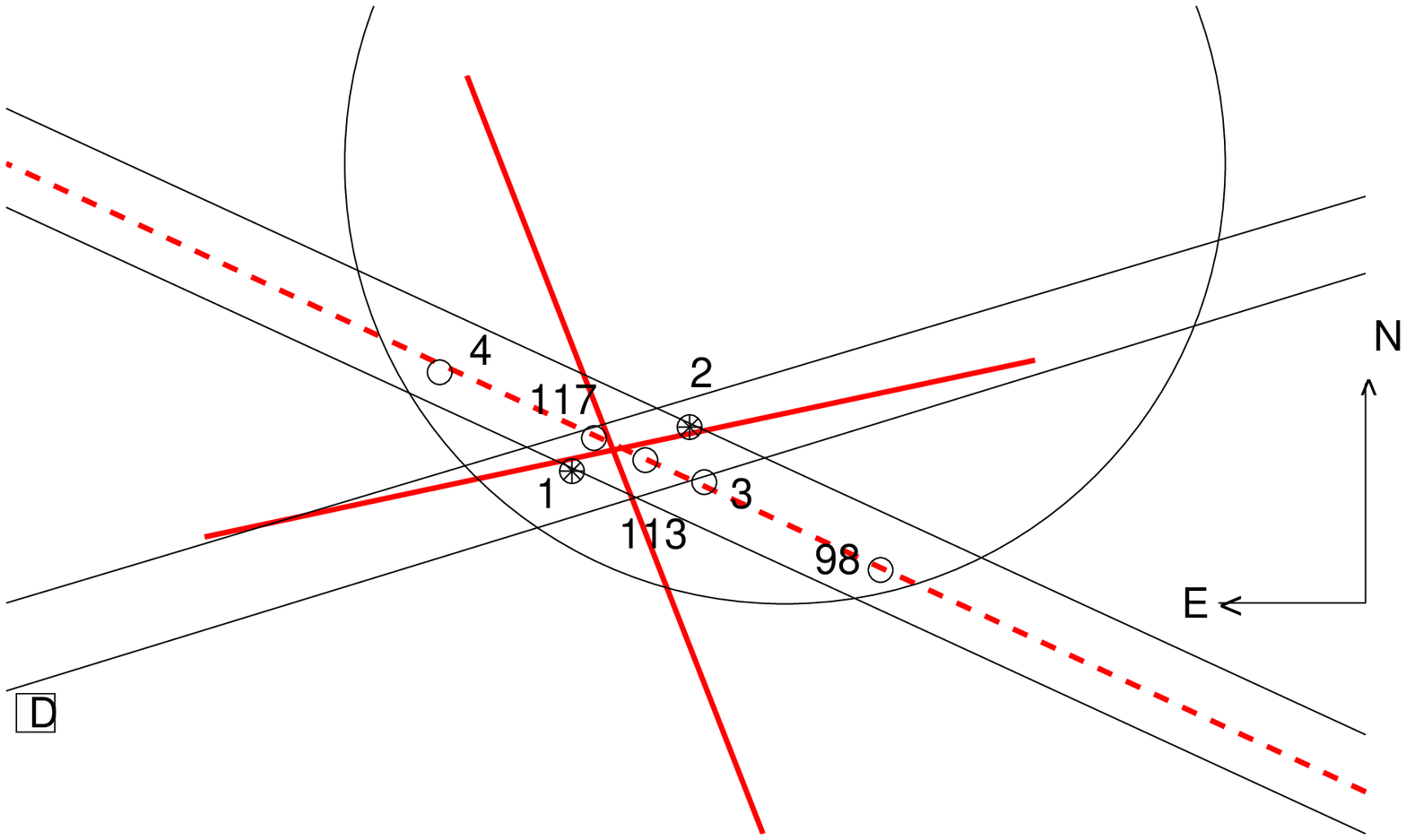}
 \caption{
Top: sketch of the slit positions and widths and of positions with strong emission lines (blobs) overlaid onto 
Gemini i-band image  adapted from NG15, fig. 2.
	 Galaxy C is roughly indicated by the white line encircled ellipse. Objects shown in pink are either foreground stars (A, B1) or galaxies (D, G, H, I). Object K could be the interacting partner of galaxy C.
	The large black circle represents the 90 percent X-ray error region.
Bottom: sketch of  slit positions and  blob locations inside the AGN photoionization cones (open circles) 
and outside the cones in the ISM (filled circles)   throughout SGRB100628A.
The red lines define the AGN photoionization cones.
} 
\end{figure*}

SGRB host galaxies  were investigated by different methods (see, e.g. Berger 2014), in particular   by  photometry 
in the infrared (IR) - radio range.
When spectroscopic  data in the optical-near IR range   were available, the  modelling  of the line ratios yielded 
valuable information about the progenitors,  in particular  by the   host metallicity trend  throughout the redshift
 (e.g. de Ugarte Postigo et al 2014, Cucchiara et al 2013, Soderberg et al 2006, Contini 2016a, 2018a  and references therein).
Moreover,  when the observations covered different positions within a single host galaxy, the results of modelling led to more
knowledge, in particular, regarding the photoionizing source of the gas and  the hydrodynamical field.
In previous papers we have  modelled in detail the line and continuum spectra of GRB host galaxies
and compared the results with  those of supernovae (SN), active galactic nuclei (AGN), starburst (SB), 
etc  at different  redshifts
(Contini 2018b and references therein) in order to   obtain the characteristic host features,
such as the gas  physical conditions and the O/H and N/H relative abundances.
For instance, the results clearly indicated that  N/O abundance ratios for LGRB decreases with the redshift
decreasing trend. However, the data  in particular  at z$>$1 for SGRB were too few to allow any conclusion.
For NGC4993, the host galaxy of the GW170817 SGRB at z=0.009873, observed by  LIGO/Virgo system through  gravitational waves,
the location within  the AGN-LINER domain in the N/O distribution diagram was clearly found (Contini, 2018a).
 The observations of SGRB  at present  do not supply enough data in order to 
decide for sure  why and when  an AGN  may appear throughout the hosts.

In this paper  we would like to  investigate the SGRB100628A host  galaxy by the detailed modelling 
 of the spectra presented by Nicuesa Guelbenzu et al (2015, hereafter NG15) and, eventually, of other SGRB hosts
on the basis  of spectroscopic and photometric data.
It is generally believed that a  star burst (SB) is the photoionizing source of the SGRB host gas,
therefore the spectra observed   from the Berger (2009) SGRB host galaxy  sample,  from the  
different positions in the SGRB100206A (Perley et al. 2012) host, from the 
various locations in the SGRB130603B host (de Ugarte Postigo et al. 2014) and from the SGRB051221a host (Soderberg et al 2006) 
were analysed by a SB dominated model. Shocks were also accounted for (Contini 2018a)  considering that SGRB galaxies derive 
from merging.
In Sect. 2 we  explain calculations and modelling of the spectra. In Sect. 3 we compare the calculated line ratios  
to those  observed by NG15 from galaxy C  and the calculated  continuum SED
to the data observed from  the SGRB100628A complex. 
The  SGRB150101B host and NGC4993 SEDs  are  discussed.
 We will show that  the interpretation of the spectra  may suggest the presence of  an AGN  
in the SGRB100628A host galaxy. Therefore,
in this paper we present new   results for  the SGRB hosts. 
In particular, N/O relative abundances calculated  on the basis of new AGN dominated models for some SGRB hosts 
are reported   in Sect. 4.  Concluding remarks follow in Sect. 5.

\section{About the calculations}

\subsection{Description of the code} 

We use composite models which account consistently for
photoionization and shocks.  The code {\sc suma} (see also Ferland et al 2016)  is adopted.
The main input parameters are those which   are used for the
calculations of the line and continuum fluxes.
They account for photoionization and heating by  primary and secondary  radiation and collisional
process due to shocks.
The input parameters such as  the shock velocity \Vs, the atomic
preshock density \n0 and the preshock
magnetic field \B0 (for  all models \B0=10$^{-4}$Gauss is adopted)
define the hydrodynamical field.
They  are  used in the calculations  of the Rankine-Hugoniot equations
  at the shock front and downstream.
They  are combined in the compression equation which is resolved
throughout each slab of the gas
in order to obtain the density profile downstream.
Primary radiation for SB in the GRB host galaxies is approximated by a black-body (bb).
  The input parameters that represent the primary radiation from the SB are the 
 effective temperature  \Ts and the ionization parameter $U$. A pure blackbody radiation 
 referring to \Ts is a poor approximation for a star burst, even adopting a dominant 
 spectral type (see Rigby \& Rieke 2004). However, it is the most suitable because 
 the line ratios that are used to indicate \Ts  also depend on metallicity, 
 electron temperature, density, ionization parameter, morphology of the ionized clouds and, 
 in particular, the hydrodynamical picture.
 For an AGN, the primary radiation is the power-law radiation
flux  from the  active centre $F$  in number of photons cm$^{-2}$ s$^{-1}$ eV$^{-1}$ at the Lyman limit
and  spectral indices  $\alpha_{UV}$=-1.5 and $\alpha_X$=-0.7. The primary radiation source
 does not depend on the host physical condition but it affects the surrounding gas.   This  region  is not considered
as a unique cloud, but as a  sequence of slabs with different thickness calculated automatically
following the temperature gradient. The secondary diffuse radiation is emitted
from the slabs of
gas heated  by the radiation flux reaching the gas and by the shock.
Primary and secondary radiation are calculated by radiation transfer.

In our model the line and continuum emitting  regions throughout the galaxy cover  an ensemble of fragmented clouds.
The geometrical thickness of the clouds is  an input parameter of the code ($D$) which is  calculated
consistently with the physical conditions and element abundances of the emitting gas.
The fractional abundances of the ions are calculated resolving the ionization equations
for each element (H, He, C, N, O, Ne, Mg, Si, S, Ar, Cl, Fe) in each ionization level.
Then, the calculated line ratios, integrated throughout the cloud thickness, are compared with the
observed ones. The calculation process is repeated
changing  the input parameters until the observed data are reproduced by the model results,  at least
within 10 percent
for the strongest line ratios and within 50 percent for the weakest ones.

However,  some parameters regarding the continuum SED, such as the dust-to-gas  ratio $d/g$  and the dust grain radius
\agr ~ are not directly constrained by fitting the line ratios.
Dust grains are heated by the primary radiation and by mutual collision with  atoms.
The intensity of dust reprocessed radiation in the infrared (IR)  depends on $d/g$ and \agr.
In this work we use $d/g$=10$^{-14}$ by number for all the models which corresponds to
4.1 10$^{-4}$ by mass for silicates (Draine \& Lee 1994).
The distribution of the grain size along the cloud starting from an initial radius
is automatically  derived by {\sc suma},
 which calculates sputtering of the grains  in the different zones downstream of the shock.
The sputtering rate depends on the gas temperature, which is $\propto$ V$_s^2$ in the immediate post-shock region.
In the high-velocity case (\Vs$\geq$ 500 \kms) the sputtering rate is so high that the grains
with \agr $\leq$0.1 \mum are rapidly destroyed downstream.
So, only grains with large radius (\agr $\geq$0.1 \mum) will survive.
On the other hand, the grains survive downstream of low-velocity shocks ($<$200\kms).
Graphite grains are more sputtered than silicate grains for T= 10$^6$ K (Draine \& Salpeter 1979).
Small grains (e.g. PAH) survive in the extended galactic regions on scales of  hundred parsecs and lead to the
characteristic features that appear in the SED.
In conclusion, cold dust or cirrus emission results from heating by the  interstellar radiation field,
warm dust is associated with star formation regions and hot dust appears around  AGN (Helou 1986) and in high velocity shock regimes.
Therefore, we will consider relatively large grains, e.g. silicate grains with an initial radius of 0.1 -1.0 \mum.

In the radio range the power-law spectrum of synchrotron radiation
created by the Fermi mechanism at the shock front is seen in most galaxies.
It is calculated by {\sc suma} adopting a  spectral index of -0.75 (Bell 1977).

\subsection{Calculation details}

The calculations initiate at the shock front where the gas is compressed and  adiabatically thermalised, reaching a maximum 
temperature in the immediate post-shock region T$\sim$ 1.5$\times 10^5$ (\Vs/100 \kms)$^2$. T decreases downstream following recombination. 
The cooling rate is calculated in each slab. The downstream region is cut  into a maximum of 300 plane-parallel slabs with 
different geometrical widths calculated automatically, to account for the temperature gradient.
In each slab, compression is calculated by the Rankine-Hugoniot equations for the conservation of mass, momentum and energy 
throughout the shock front. Compression (n/\n0) downstream ranges between 4 (the adiabatic jump) and $\geq$10, depending on \Vs and \B0. 
The stronger the magnetic field, the lower the compression downstream, while a higher shock velocity corresponds to a higher compression.
The ionizing radiation from an external source is characterised by its spectrum and by the flux intensity. 
The flux is calculated at 440 energies from a few eV to keV. Owing to radiative transfer, the spectrum changes throughout 
the downstream slabs, each of them contributing to the optical depth. In addition to the radiation from the primary source, 
the effect of the diffuse radiation created by the gas line and continuum emission is also taken into account, using 240 energies 
to calculate the spectrum.  For each slab of gas, the fractional abundance of the ions of each chemical element is obtained by solving the ionization equations. 
These equations account for the ionization mechanisms (photoionization by the primary and diffuse radiation, and collisional
 ionization) and recombination mechanisms (radiative, dielectronic recombinations), as well as charge transfer effects. 
The ionization equations are coupled to the energy equation if collision processes dominate, and to the thermal balance if 
radiative processes dominate. The latter balances the heating of the gas due to the primary and diffuse radiations reaching the 
slab with the cooling due to recombinations and collisional excitation of the ions followed by line emission, dust collisional 
ionization and thermal bremsstrahlung. The coupled equations are solved for each slab, providing the physical conditions necessary 
for calculating the slab optical depth, as well as its line and continuum emissions. 
The slab contributions are integrated throughout the cloud.  
In particular, the absolute line fluxes corresponding to the ionization level i of element K are calculated 
by the term nK(i), which represents the density of the ion i. 
We consider that nK(i) = X(i) [K/H] n$_H$, where X(i) is the fractional abundance of the ion i 
calculated by the ionization equations, [K/H] is the relative abundance of the element K to H and n$_H$ 
is the density of H (in number cm$^{-3}$). In models including shock, n$_H$ is calculated by the compression equation (Cox 1978) 
in each slab downstream. Accordingly, the abundances of the elements are given relative to H as input parameters.

The N$^+$ and O$^+$ ions are linked by charge exchange reactions 
with H$^+$, so they follow the same profile throughout the  clouds. 
However, the [NII]/\Hb line ratios are observed  for blobs 1, 3, 4 and 98,
indicating  that  the N/H relative abundance  varies sensibly from blob to blob.
 Changing  the element abundances  (e.g. the N/H ratio) by a large factor ($>2$) in a model
will  affect the cooling rate
in the downstream recombination region throughout a cloud of gas. Consequently the calculated  line ratios
will change and the modelling process  will be restarted with different input parameters
in order to recover a good fit to all the observed line ratios.

Dust grains are coupled to the gas across the shock front by the magnetic field (Viegas \& Contini 1994). 
They are heated by radiation from the AGN and collisionally by the gas to a maximum temperature, which 
is a function of the shock velocity, of the chemical composition and of the radius of the grains, 
up to the evaporation temperature (T(dust) $\geq$ 1500 K). The grain radius distribution downstream 
is determined by sputtering, which depends on the shock velocity and on the density. Throughout shock fronts 
and downstream, the grains might be destroyed by sputtering.

Summarizing, the code starts by adopting an initial \Te (10$^4$ K) and the input parameters for the first slab. 
It then calculates the density from the compression equation, the fractional abundances of the ions 
from each level for each element, 
line emission, free-free emission and free-bound emission. It re-calculates \Te by thermal balancing or the enthalpy equation, 
and calculates the optical depth of the slab and the primary and secondary fluxes. 
Finally, it adopts the parameters found in slab i 
as initial conditions for slab i + 1. 
Line and continuum intensities are integrated  accounting for all the  slabs.  The number of lines  calculated by 
{\sc suma}  is  $>$ 200 for each model.
 They are calculated  at the gaseous nebula that emits the spectrum, while
 the data are observed at Earth. Therefore they diverge by a factor (r$^2$/d$^2$) that depends on the distance of the 
nebula from the  radiation centre (r), 
and on the distance (d) of the galaxy to Earth. We then calculate the line ratios to a specific 
line (in the present case \Hb, 
which is a strong line), and compare them with the observed line ratios.

 On this basis we calculate a grid of models. The  set of  models   (e.g. Table 1)
which best reproduce the line ratios is selected. We  obtain the final model by cross-checking
the fit of the calculated continuum  SED  to the observed one.

\subsection{Modelling the continuum SED}

   The  models constrained by the  line spectra
   give a hint about the relative importance of the different
ionization and heating mechanisms which  are recognised throughout the continuum SED
in each of the objects. 

The gas ionized by the  SB (or AGN) radiation flux  emits continuum radiation (as well as the line fluxes)
from  radio to  X-ray.
The continuum accounts for
  free-free and free-bound radiation (hereafter  addressed
to as bremsstrahlung).
The bremsstrahlung  at $\nu$$<$ 10$^{14}$ Hz has a similar slope in
all the diagrams.
In fact, the
bremsstrahlung continuum, emitted by free electrons accelerated in
Coulomb collisions with positive ions
(mostly H$^+$, He$^+$ and He$^{++}$) in nebulae of charge Z has an
emission coefficient (Osterbrock 1974):

J$_{\nu}$ $\propto$\Ne N$_+$Z$^2$($\pi h\nu$/3kT)$^{1/2}$ e$^{-(h\nu/kT)}$    (1)

\noindent
The  photoionization radiation flux can heat the gas to T$\sim$ 2-4$\times$ 10$^4$ K, while the
gas is heated  collisionally by
the shock to a maximum of T$\propto$ (\Vs)$^2$, where
\Vs is the shock velocity.
The cooling rate downstream  depends on  \Ne N$_+$ (N$_+$ is the proton density).
The trend of the bremsstrahlung as function of $\nu$  follows the
interplay between T  and $\nu$.
High temperatures of the emitting gas  determine the maximum bremsstrahlung at high $\nu$.
At T$\sim$1-4$\times$10$^4$ K the exponential term is significant
at frequencies between  10$^{14}$ and 10$^{15}$ Hz.
The  temperatures  are calculated by thermal balancing between the heating rates which depend on the
photoionizing flux and  the cooling rates by free-free, free-bound and line emission.
Therefore,  the radiation effect is seen mainly in this frequency  range.
In the radio range, the exponent  in eq (1)
tends to 0 and  the continuum is $\propto \nu^{1/2}$.
So the SEDs in all the  diagrams  of all  galaxy types have  similar trends  at
relatively low frequencies and  the dust reprocessed radiation bump in the IR is clearly recognizable.

In conclusions,
i) the black body radiation corresponding directly to the temperature
dominating in the starburst  is seldom observed  in the
UV, because  absorption  is very strong in this frequency range
due  to  strong line formation.
ii) The shock effect throughout the SED can be recognized from  the maximum frequency and
intensity of the dust reprocessed radiation peak
in the infrared and of the bremsstrahlung   at high frequencies.
iii) 
The gas ionized by the  SB (or AGN) radiation flux  emits  bremsstrahlung
from  radio to  X-ray.
The black body emission from the background old star  population with
T$_{bb}$$\sim$ 3000-8000 K  generally emerges
over  the bremsstrahlung  throughout the   SED in the near-IR(NIR) - optical range.
iv) In the radio range synchrotron radiation created by the Fermi
mechanism  is recognized by its spectral index.
   Thermal bremsstrahlung in the radio range has a steeper trend which becomes even steeper
by self-absorption at low $\nu$.
In the far-IR only comparison with the observation data  indicates the source of the continuum
radiation flux, because  thermal bremsstrahlung, synchrotron radio and cold dust reradiation may be
blended.

\section{Analysis of  SGRB100628A  spectra}

 The GRB100628A burst triggered the Swift/Burst Alert Telescope (BAT)(Barthelmy et al 2005).
In the BAT energy window (15-350 keV) the burst had a duration
of T$_{90}$ (0.036$\pm$0.009s, Immler et al 2010).
 NG15  used the multi-colour imager  Gamma-Ray Burst Optical/Near-Infrared Detector (GROND) that is an imaging instrument
at the European Southern Observatory (ESO) in La Silla MPG 2.2 m
telescope, ESO/Very Large telescope (VLT) spectroscopy, and deep Australia Telescope Compact Array (ATCA) radio-continuum observations
together with publicly available Gemini imaging data to  investigate the  suitable host and the galaxies in the field
of GRB100628A.

NG15 searched for the SGRB100628A host  among  galaxies within and outside the XRT  circle.
 They reported that the  burst was followed by a faint X-ray afterglow but no optical 
counterpart was discovered.
However, inside the  XRT circle "a potential host galaxy at a redshift of z = 0.102 was soon  observed" as
a morphologically disturbed, interacting galaxy system C at z = 0.102. 
NG15 claim that the objects within the XRT circle  are connected by
tidal streams reaching several kpc distances.
We suggest that the host morphological structure is  most likely shaped by  merging of the  progenitors.
Strong  emission lines of
[OII], [OIII], \Ha and \Hb  were revealed by VLT/FOcal Reducer and
low dispersion Spectrograph (FORS2) spectroscopy
indicating  star formation activity and suggesting that the  progenitors  belonged to a
young stellar population. 
Another galaxy (galaxy D),
 a radio-bright, luminous elliptical galaxy at a redshift z = 0.311  was
observed  outside the XRT. 
 NG15 claim that the "higher redshift solution for galaxy D  fits slightly better 
into the ensemble properties of short GRBs known so far."
The  redshift (z=0.311) of galaxy D is  close to those found for
other SGRB (z$\sim$ 0.4). This  argument may however change  by  future observations.

\subsection{Galaxy C. Line ratios}

\begin{table*}
\centering
\caption{Modelling  the corrected observed line ratios to \Hb in the different blobs  of SGRB100628A}
\begin{tabular}{lcccccccccccccccc} \hline \hline
\ line      & [NeIII]  & [OII]  & \Hg  & \Hb & [OIII]  &  \Ha &  [NII]   & \Hb$_{obs}^1$ &  \Hb$_{calc}^2$ \\ 
\           & 3967     & 3727+  & 4340 &4861 & 5007+   & 6563 &  6583    & 4861          &   4861           \\ \hline
\ OBS$_1$   &   -      & 7.15   & 0.36 &  1  & 4.13    &  3   &  0.96    & 1.9           &   -              \\
\ mod$_1$   &   -      & 7.2    &0.45  &  1  & 4.1     &  3   &  1.      &  -            & 0.002            \\
\ mod$_1^*$ &   -      & 6.9    &0.45  &  1  & 4.17    &  3   & 1.1      &  -            & 6.6e-4           \\  
\ OBS$_2$   & 0.7      &3.95    &0.8   &  1  &4.98     &  3   &  -       &  2.8          & -                \\
\ mod$_2$   & 0.54     &3.8     &0.46  &  1  &4.94     & 2.95 & -        &    -          & 0.0022           \\  
\ mod$_2^*$ & 0.6      &3.98    & 0.45 &  1  &4.99    & 3.07 & -        &   -           & 5.5e-4           \\ 
\ OBS$_3$   &  -       &-       &  0.38&  1  & 5.14    & 3    & 0.31     & 6.2           & -                 \\
\ mod$_3$   &  -       & 0.73   & 0.46 &  1  & 5.2     & 2.95 & 0.36     &   -           & 0.084        \\
\ mod$_3^*$ & -        & 0.88   & 0.46 &  1  & 5.05    & 3.08 & 0.5      &   -           & 8.87             \\ 
\ OBS$_4$   &  -       & -      & -    &  1  & 6.6     & 3    & 1.4      &  1.4          &  -               \\
\ mod$_4$   &  -       &0.77    &0.46  & 1   & 6.65    & 2.95 & 1.2      &   -           & 0.08             \\
\ mod$_4^*$ &  -       &1.4     &0.46  & 1   & 6.7     & 3.   & 1.2      &  -           & 0.69             \\   
\ OBS$_{98}$ &  0.15   & -      & -    & 1   & 4.09    & 3    & 0.07     &  1.2          & -                 \\
\ mod$_{98}$ & 0.3     & 0.2    & 0.46 & 1   & 4.04    & 3    & 0.07     &   -           &25.14             \\
\ mod$_{98}^*$ & 0.4   & 0.8    & 0.46 & 1   & 3.9     & 3    & 0.1      &   -           &11.74             \\  
\ OBS$_{113}$  & -     & -      & -    & 1   & 5.88    & 3    &  -       &  2.5          & -                \\
\ mod$_{113}$  & -     & 0.7    & 0.46 &  1  & 5.86    & 3    & 0.4      & -             & 0.092            \\
\ mod$_{113}^*$ & -    & 0.9    & 0.46 & 1   & 5.89    & 3    & 0.5      &  -            & 6.9              \\ 
\ OBS$_{117}$  & -     & -      & -    & 1   & 4.72    & 3    &  -       &  0.43         & -                \\
\ mod$_{117}$  & -     & 0.7    & 0.46 & 1   & 4.75    & 3    & 0.064    & -             & 0.1              \\
\ mod$_{117}^*$ & -    & 0.7    & 0.46 & 1   & 4.75    & 3    & 0.8      &  -            & 13.3             \\   \hline

\end{tabular}

$^*$ AGN dominated models;  $^1$  line flux in 10$^{-17}$\erg observed at Earth; 
$^2$  line flux in  \erg  calculated at the nebula.

\end{table*}

\begin{table*}
\centering
\caption{SB and AGN dominated models  for  the different blobs}
\begin{tabular}{lcccccccccccccccc} \hline \hline
\           & \Vs   & \n0   &  $D$        &  \Ts    &   $U$  & $F$      & log(O/H)+12   &  log(N/H)+12  \\  
\           & \kms  & \cm3  & 10$^{18}$cm & 10$^4$K &    -   & $^1$     &     -         &       -       \\ \hline
\ mod$_1$   & 120   & 70    & 0.08        & 8.8     & 0.0033 &   -      &   8.82        & 7.6            \\
\ mod$_1^*$ & 120   & 70    & 0.025       &   -     &    -   & 0.037    &    8.82       & 7.7            \\ 
\ mod$_2$   & 90    & 70    & 0.014       & 7.2     & 0.0043 & -        &   8.82        & 7.7            \\
\ mod$_2^*$ & 110   & 60    & 0.035       &   -     &    -   & 0.042    &   8.83        & 7.7            \\ 
\ mod$_3$    &100    &100    & 15          & 6.0     & 0.1    & -       &   8.82        & 8.0         \\  
\ mod$_3^*$  &300    &300    & 2.9        &   -      &   -   & 100      &    8.82        & 7.7         \\ 
\ mod$_4$    &100    &100    &  9          & 6.9     & 0.1    & -       &   8.82        & 8.3          \\  
\ mod$_4^*$  &200    &400    & 5.0         &  -      &  -     &10       &   8.82        & 7.9          \\  
\ mod$_{98}$ &300    &500    & 1.3         & 10.     & 10.    & -       &   8.82        & 7.3          \\
\ mod$_{98}^*$ &300  &300    &4.2          &  -      &   -    &96       &   8.82        & 7.0          \\ 
\ mod$_{113}$& 100   &100    &15           &  6.8    &  0.1   &  -      &   8.82        & 7.78         \\
\ mod$_{113}^*$ &300 &300    &2.2          &  -      &   -    &100      &   8.82        & 7.48         \\ 
\ mod$_{117}$&  100  &100    &15           &  6.4    &  0.1   &  -      &   8.82        & 7.0          \\
\ mod$_{117}^*$ &300 &300    &5.0          &  -      &   -    &100      &   8.82        & 7.7          \\ \hline

\end{tabular}

$^1$  in  10$^{10}$ photons cm$^{-2}$ s$^{-1}$ eV$^{-1}$ at the Lyman limit;

\end{table*}

 In  Fig. 1  we report VLT/FORS2 long-slit spectroscopy presented by NG15  (in their fig. 2)   
of the GRB100628A field 
obtained  by two slit positions covering the nine blobs (defined as  positions detected with strong emission 
lines in the 2D spectra) 1, 2, 3, 4, 94, 98, 113, 117, 136.
Numbers larger than 4 refer to the pixel coordinate along the corresponding slit.

 All the observed lines are at  the  common redshift z$\sim$0.102$\pm$0.001.
 We have corrected  the  data  presented by NG15 (in their table 3)
following  Osterbrock (1974) equations for the Milky Way.
 Cross-checking the dust-to-gas (d/g) ratio which derives from reproducing the observed IR data by dust 
 reprocessed radiation flux  from the
GRB100628A host (Sect. 3.2), we have  found that the WISE satellite data (which are all upper limits)
constrain d/g to $<$ 0.0004 by mass. This value has been adopted by the models (Sect. 2.1). 
The average d/g in the Milky Way is 0.007 (Dwek \& Cherchneff 2011).
Therefore the reddening correction for dust obscuration is  mainly Galactic.
We adopted \Ha/\Hb=3  at the nebula because in the presence
of shocks the   \Ha/\Hb  ratios  range  between  3.05 and 2.87
considering an emitting gas with T  from $<$5000 K  to $>$10000 K.
\Ha/\Hb $\geq$ 4 can be found for  high optical depths (Osterbrock 1974, fig 4.3).

We analysed the observed spectra by comparing  calculated line ratios  to the observed ones in Table 1
 (in order to avoid the effect of distances). 
The spectra indicated as OBS$_1$ - OBS$_{117}$  correspond to  the observed  
line ratios to \Hb for blobs 1 to 117 (columns 2-8), respectively. The last two columns show 
the \Hb line flux observed
at Earth  (in units of 10$^{-17}$ \erg) and the \Hb line fluxes calculated at the nebula (in \erg),
respectively, for each model.
 We define a model as the set of the input parameters which  yields the best fit to the data.
In  composite models  the  number of the significant input parameters
is relatively high (see Sect. 2)  because it accounts for both the photoionization  and the shock 
(see Sect. 2) but it does not depend on the number of  the observed lines.
For each model the code calculates the flux of the most significant lines (more than 200) from the far- UV to the far -IR.
The strong  lines of a relatively abundant element (e.g. [OIII]5007+ and [OII] 3727+) 
from  at least two ionization levels and the \Ha and \Hb lines  constrain the model.
 Table 1 shows that although the [OIII] lines  were observed in all the reported blobs,
the [OII] lines were observed only in blobs 1 and 2. 
NG15 report three runs of VLT/FORS2 observations. In the third run (on 2013 April 03)   they used the 
GG 435 order separation filter  which does not cover the [OII]3727+ line wavelengths.
because it  cuts off the lines at $\lambda$$<$ 4400\AA.
Then,  the observed [OII] lines are not shown in blobs   3, 4, 98,  113 and 117 spectra, while 
they are relatively strong in  those corresponding to blobs 1 and 2.
The [OII] lines together with the [OIII]5007+ lines are basic to constrain the models.
We wonder why  they were avoided in the  observations of the blobs  which are located not far from blobs 1 and 2.
We consider two cases. First, the [OII] line fluxes  were of the same order  as those observed in blobs 1 and 2. Then
we can predict, qualitatively,  that the physical conditions  throughout galaxy C are  almost homogeneous.
As a second choice  we assume that the  [OII] line  fluxes were   very low  and the lines were
blended  with  the noise  (see NG15,  their fig. 4). 
The [OII] line weakness  is   used to constrain the models and we can proceed to model the line ratios.
We   report in Table 1 the calculated [OII]/\Hb line ratios  
even if the   [OII] lines were not observed.  
This  procedure is valid only if the  [OII] lines were
missing in the spectra because very weak. 

A SB is generally adopted to explain the photoionizing flux in GRB hosts
by the observer community.
To have a first hint about the initial parameters to be used in  modelling, we have used the grids
 calculated by the code {\sc suma} for models
which account for both photoionization and shock in  SB and AGN cases (Contini \& Viegas 2001a,b).
Then,  refined models more  suitable to each spectrum were calculated.
Modelling results  in terms of the selected physical parameters and the relative
abundances (SB models mod$_1$   to mod$_{117}$)  are  shown in Tables 1 and 2. 
 The line observed fluxes by NG15 do not include  the errors, therefore we tried to reproduce them
as close as possible. Table 1  and  Figs. 2 show that the errors are $<$ 4 percent.
In  Table 2 columns 2-4 the shock velocity, the preshock density and
the cloud geometrical thickness    describe the  shock. Columns 5 and 6 show the parameters  for the SB
(the  effective temperature and the  ionization parameter, respectively).
In columns 8 and 9, O/H and N/H relative abundances calculated by the detailed modelling  of the
line ratios are reported.

 Table 1 shows that the  [OIII]/\Hb line ratios range between a maximum of 6.6 in blob 4
and a minimum of 4.72 in blob 117.  [OII]/\Hb line ratios  in blobs 1 and 2   have
relatively high values of 7.15 and 3.95, respectively.
This indicates that   blob 2 and blob 3 spectra,  although  showing  similar [OIII]/\Hb  ($\sim$5), 
are emitted  from gas in different physical conditions.
 The same is valid  for blobs 1  and 117. 

The best results calculated adopting SB models  show that in blobs 1, 2, 3, 4, 113 and 117,
the shock velocities  are relatively low (\Vs $\leq$120 \kms) typical of clumps in SGRB host galaxies (Contini 2018a).
Preshock densities (\n0 $\sim$ 70-100 \cm3) are lower by a factor of 2 than in other SGRB hosts.
In blob 98, \Vs and \n0 are relatively high
(300 \kms and  500\cm3, respectively).
The geometrical thickness $D$ of  blobs 1 and 2 clouds (Table 2)
are smaller by factors of 100-1000 than for all the other blobs.
$D$ ranges between a minimum of$\sim$0.0046 pc for blob 2 to
a maximum of $\sim$5 pc for blob 117. 
Cloud fragmentation  ($\propto 1/D$) is  very high   in blob 2,
lower by a factor of $\sim$6
in blob 1 and very low in blobs 3, 4, 98  and 113.
Moreover, for all the models the calculated $D$  are lower by a large factor  than the slit widths
 adopted by NG15 observations  (1".0 - 1".3 which correspond at the GRB100628A distance of 465 Mpc 
to $\sim$2 kpc.)
This suggests that  a relatively large number of  clouds may contribute 
to each of the observed spectra.
To model the  spectra of blobs leaking the [OII] lines,  SB temperatures $<$ 7$\times$10$^4$ K (except for
blob 98) were adopted
and $U$ higher by factors  of $\sim$ 25-30 than for blobs 1 and 2. Low [OIII]/[OII] line ratios could not be fitted by  
\Ts $<$ 7$\times$10$^4$K and a high  $U$ (see Contini 2016a, fig. 1).

So far we have been  dealing with   the SB throughout galaxy C. 
 The physical conditions calculated by  SB  models for the nine  blobs aligned within the slit  may show the  
trace of  merging. 
It was  suggested that the zones of higher pre-shock densities in merger galaxies (e.g. NGC 3393 Contini 2012)
show the collision regions.
Blobs 1 and 2   with lower densities and a lower ionization parameter are  more adapted to represent the ISM.   
Concerning the radiation source, \Ts and $U$ show jumps which  are unusual from different blobs at small  distances.
Therefore, we have tried to  reproduce the spectra by a different radiation type: a power-law flux. 
 We have  found  a  suitable fit to the observed  line ratios 
also for  the power-law dominated models mod$_1^*$  to mod$_{117}^*$.
Table 1  shows that the data are well reproduced adopting
both  black-body and  power-law dominated models.
Power-law photoionization fluxes are  used to  reproduce the line ratios emitted from  AGN galaxies. 
The results of  AGN  models (Table 2)  show \Vs$\geq$ 200 \kms and \n0$\geq$ 300 \cm3 for the clouds  in all the blobs, 
except for blobs 1 and 2. 
Moreover, log($F$)$\geq$11 in all the blobs, except for 1 and 2, where $F$ is lower by a factor $>$ 200.
The flux $F$ from the AGN is given in Table 2, column 7.
 The precision of model calculations is cross-checked with the observations in Fig. 2.
 A more  suitable  explanation of the parameter distribution  throughout the GRB100628 host galaxy 
is suggested by the  power-law model results.
It  allows to draw the  ionization cones of  an  AGN (Fig. 1) which  is not seen but  emerges from the 
interpretation of the  spectra. 
Blobs 1 and 2 are located in the ISM  regions external to the AGN cones.
Two or more  photoionization sources can coexist in  GRB hosts
and in  merger systems (Contini 2012 and references therein).
 It was found  e.g.  for a sample of AGN galaxies
at z$\sim$0.8 (Contini 2016b) that half of them show multiple radiation sources, a SB and
an accreting AGN.

Moreover, AGN in SGRB are not unusual.
An AGN dominates 
in NGC4993, the host galaxy of GW170817 (Levan et al. 2017, Contini 2018a). 
 In particular, it was  found that the bremsstrahlung emitted by the  AGN clouds can be seen throughout 
the SED in the radio and in the UV - X-ray frequency domain.
 NG15 and Berger (2010) reported that the first revised error circle (Starling et al 2010) of SGRB100628A
  enclosed an AGN and  
Xie et al. (2016) found an AGN in the host galaxy of GRB150101B.
In conclusion,  to model SGRB100628A spectra, we  adopt two model types,  
one dominated by a black body radiation flux
for the SB  and  the other dominated by  a power-law flux
for the AGN. Both  models account also for the shocks.

\begin{figure}
\centering
\includegraphics[width=8.0cm]{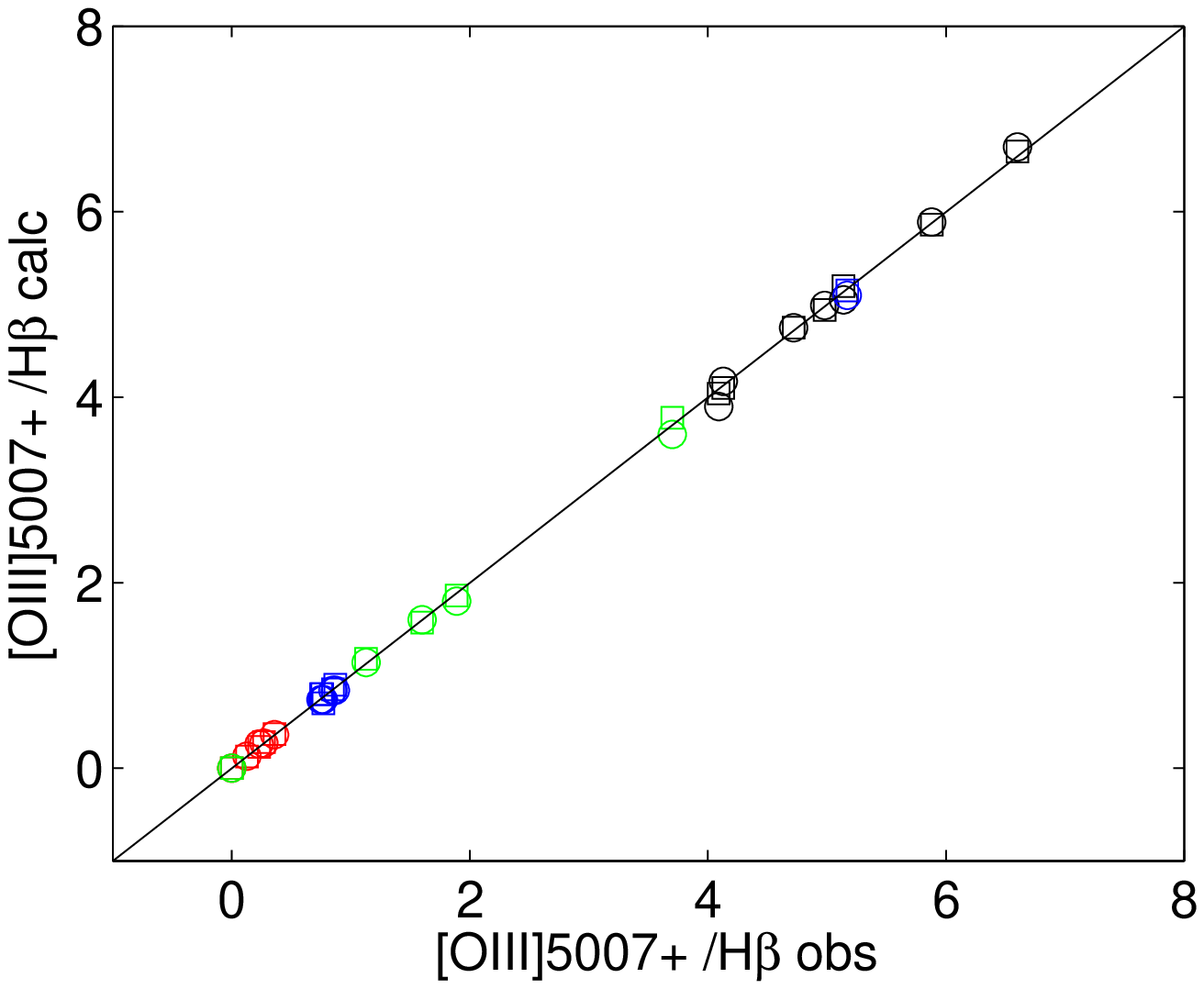}
\includegraphics[width=8.0cm]{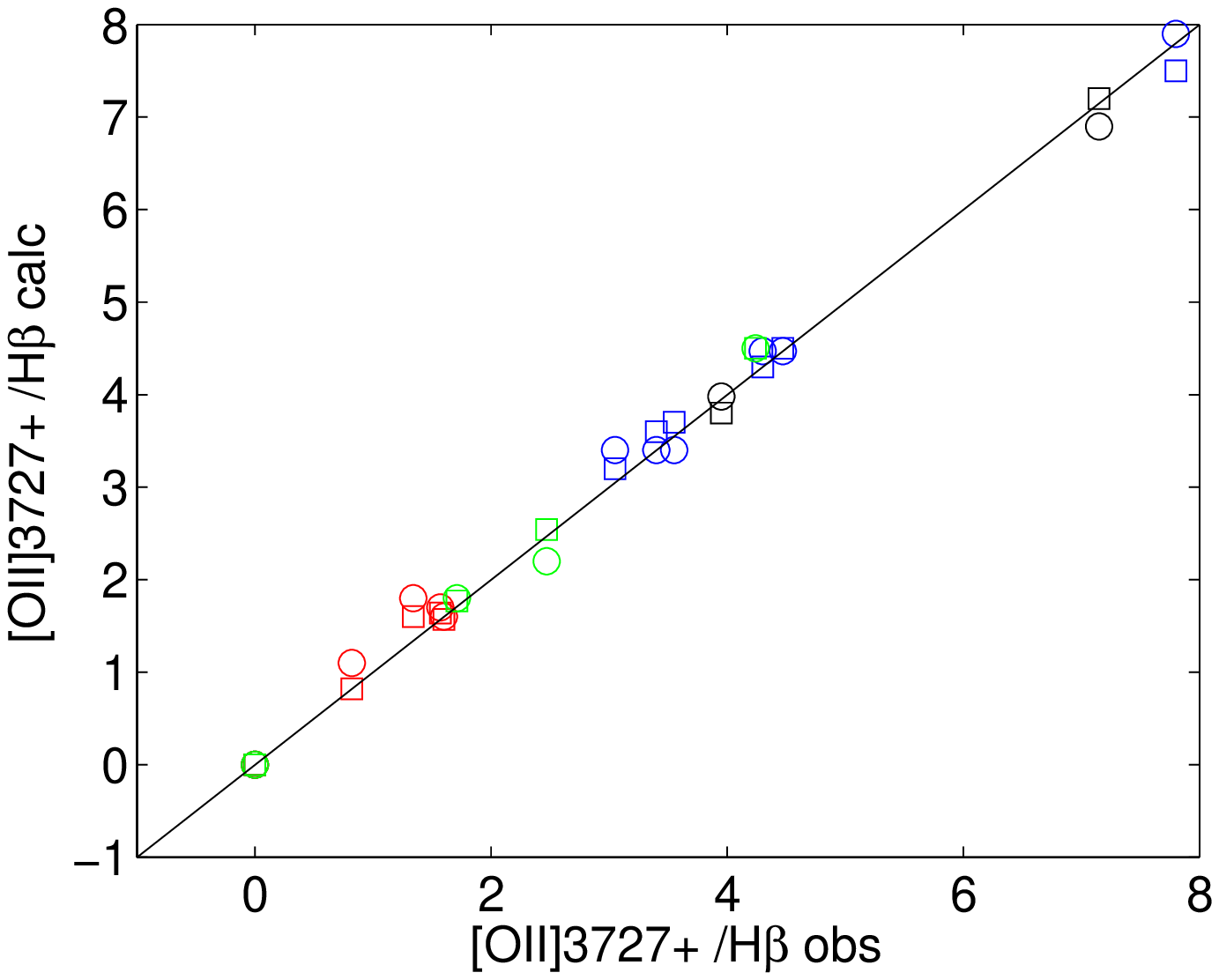}
\includegraphics[width=8.0cm]{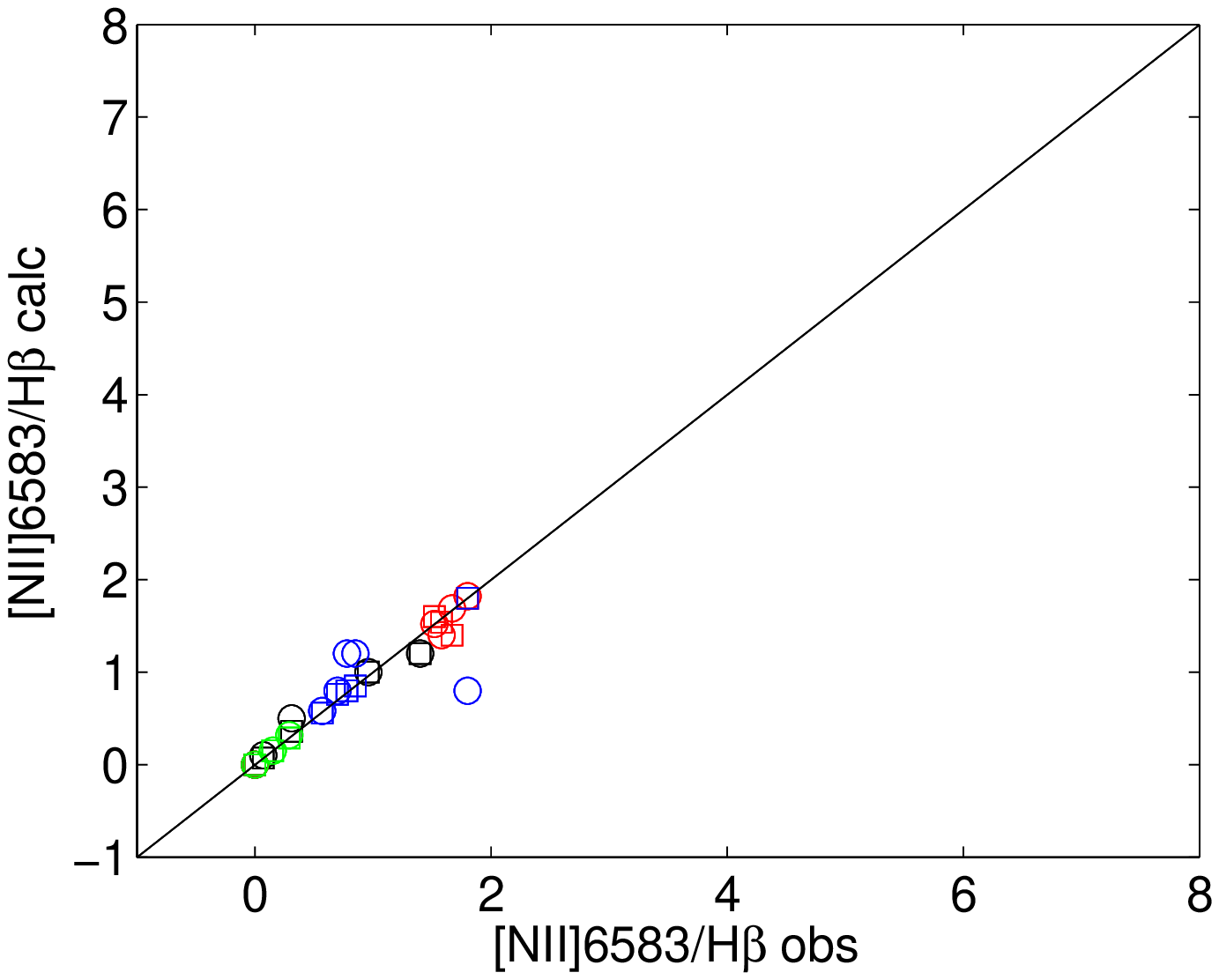}
 \caption{
Cross-checking calculation precision  of the calculated  versus the observed line ratios:
[OIII]/\Hb (top), [OII]/\Hb (middle), [NII]/\Hb (bottom). Squares: SB models; circles: AGN models.
Black: NG15 data; red: Perley et al (2012) data; blue: data  from de Ugarte Postigo et al (2014),
Cucchiara et al (2013), Soderberg et al (2006); green: data from the the Berger  (2009) sample (Sect. 4).
}
\end{figure}

\begin{figure}
\includegraphics[width=9.4cm]{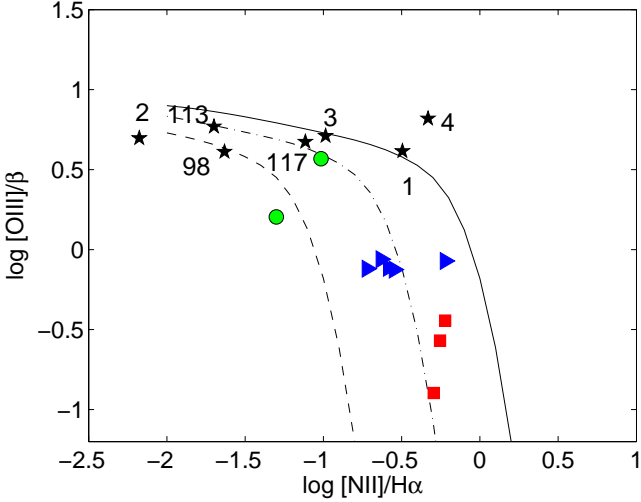}
\caption{Comparison of data with  BPT diagram. Solid line: solar N/H; dot-dashed line: N/H= 1/3 solar;
dashed line: N/H= 1/10 solar. Black stars: NG15; red squares: Perley et al (2012); blue triangles:
de Ugarte Postigo et al (2014)
and other SGRB hosts; green circles: Berger (2009)}
\end{figure}

 The results   may be further checked  by the  other methods described in the following.
 It was suggested by Contini (2016b and references therein) that the [OIII]5007+/[OIII]4363 line ratios 
could be used to
distinguish  AGN   from SB dominated spectra. However, the [OIII] 4363 line  is not measurable in the
NG15 spectra. 
Alternatively, (Baldwin, Phillips \& Terlevich 1981)  BPT diagrams are generally used 
to separate  SB from  AGN by cross-checking the  line ratios.   
 We can only  make use of
the diagram  corresponding to [OIII]/\Hb versus [NII]/\Ha, because the [SII] and  the [OI] lines are not reported
by  NG15. 
The  spectra which show both [OIII]/\Hb and [NII]/\Ha line ratios  
are few because  [NII] line fluxes are given  only for blobs 1, 3, 4 and 98. They are   reported in  Fig. 3. 
For blobs 2 , 113 and 117 we have used as an upper limit the minimum [NII] line flux
observed from the blob ensemble. The [NII]/\Ha
line ratios have been reddening corrected for each blob.
N/H relative abundances in GRB hosts are often lower
than  solar,  therefore the  data  can be  misinterpreted  through the  original BPT diagram (cf Kewley et al 2001).
Therefore we have roughly corrected the BPT diagrams by reducing the [NII]/\Ha line ratios.
Fig. 3 shows that blob 4 is in the AGN domain  above and at the right of the separation lines
in the diagram.
 Blobs 1 and 3  are on the border-line, blobs 2 and 98 are in the SB domain close  to
the border-lines for depleted N/H, while blobs 113 and 117 are
 located   close to the  AGN region  with slightly depleted N. 
The interpretation of the 113 and 117 spectra is very uncertain because
 the [OII] and [NII] lines were not observed.
The information obtained from the BPT diagram is, however,   uncertain because  the error bars are missing.

In order  to determine whether the SB or the AGN dominates in each blob,
 we   suggest a different criterion.
We have selected the  dominant photoionization source (a SB or an AGN) in each blob   by comparing 
the  \Hb flux calculated by a SB dominated model  to  \Hb calculated by an AGN (Table 1).  
We have found by the detailed modelling method that the AGN  prevails in  blobs 3, 4, 113 and 117  with  
a  maximum \Hb flux in blob 117.
 The  relatively high  \Hb flux,  which is   appropriate to an AGN, 
decreases by a factor  $<$2 in blobs 3 and 113 and by a factor $<$ 20 in blob 4.
The \Hb calculated flux for the AGN is high in blob 98  but lower than   \Hb calculated by the SB model.
The AGN dominated blobs are aligned   throughout the  North-East/South-West  slit direction  (Fig. 1). 
We suggest that the alignment of blobs 4, 117, 113, 3 and 98 indicates the axis of the photoionization cones in the AGN
with apex  most likely situated  between blobs  117 and 3 (where $F$ is  maximum). The cone  maximum opening angle
is constrained by the location of blobs 1 and 2.  The  angle is  close to those predicted by Wilson (1996) 
and Keel et al (2018), who claim that 
ionization cones have  projected opening angles of typically 70$^o$. 
Shock velocities of 200-300 \kms and preshock densities of 300-400 \cm3  which are characteristic of the AGN narrow line region (NLR),
  confirm this hypothesis.
On the other hand, shock velocities and
preshock densities of the gas inside blobs 1 and 2 are  suitable to  the ISM,  outside the cones.  
In blobs 1 and 2,  \Hb calculated fluxes for SB models  are higher than for AGN models. 
Moreover, the physical conditions are similar to those found in SB dominated host galaxies.
In  blobs 1 and 2, $F$ is lower. 
A low power-law flux (log$F$$<$10)  cannot be neglected  because  it is  similar
to   the flux generally  calculated in  the outer  NLR of AGN (e.g. Contini  et al 2002).
A low  $F$, adapted to a low luminosity AGN (LLAGN) or to a low ionization narrow emission region (LINER),
was  found   in NGC4993, the host galaxy of GW170817 (Contini 2018a). 
Moreover, the observed [OII]/\Hb and [OIII]/\Hb line ratios  in blobs 1 and 2 are similar to those
presented by Ho et al (1993) in their sample of LINERs for galaxies NGC1167 and NGC1275. (However, the
[NII]/\Hb ratios are higher by a factor $\leq$5 for the two LINERs than for galaxy C.) 
These objects  were classified  as a LINER and a
shock dominated galaxy, respectively, by Contini (1997).

Blob 98 which is  located  in  the  tidal tail of galaxy C   at a distance of 7" (which is translated to
13 kpc at  z=0.102) deserves  particular attention. 
  This distance is roughly reached  by  the   photoionization cone   in the  AGN extended NLR.
AGN characterised by conical or biconical structures  were observed up to  $\geq$ kpc distances and the extended NLR 
even to 12-20 kpc  (Mulchaey et al 1996, Schmitt et al 2003).
We have found that the \Hb flux emitted by the clouds photoionised by the SB  overcomes by only a factor of 
$\sim$ 2  the AGN \Hb flux.
Moreover, the emitting gas shows  in blob 98 \Vs and \n0 higher than those characteristic of GRB hosts (Contini 2016a).
The high $U$  in blob 98  indicates that the  emitting clouds are close to the black-body radiation source.
Therefore, we  suggest that the SB  bulk  is  located  in
the tail of galaxy C   which  is also reached by the  AGN flux.  
In conclusion,  the clouds in blob 98 show the signature of both the SB and the AGN.
Blob 3, 4,  113 and 117 spectra throughout SGRB100628A are more properly reproduced by  AGN models.

For all the gaseous clouds within the blobs,  O/H relative abundances are solar 
((O/H)$_{\odot}$=6.6 $\times$ 10$^{-4}$, (N/H)$_{\odot}$= 10$^{-4}$, 
(S/H)$_{\odot}$= 1.6 $\times$ 10$^{-5}$ and (Ne/H)$_{\odot}$= 10$^{-4}$, Grevesse \& Sauval 1998).  
Ne/H are lower than solar in blob 98.  12+log(N/H)  spans  between a maximum of 8.3  
and a minimum of 7 and S/H changes from blob to blob depending, in particular,  on the S  atomic fraction  
trapped into dust grains.

\subsection{Fitting the  observed continuum SED}

\begin{table}
\caption{Description of the data for GRB100628A SED in Fig. 4}
\begin{tabular}{lccccccccccc} \hline \hline
\
\  Inside  the XRT   &     &      \\
\ W1(3.4\mum)$^1$ & $>$20.0 AB mag. & black triangle    \\
\ W2(4.6\mum)$^1$ & $>$20.2 AB mag. & black triangle     \\
\ W3(12\mum)$^1$ & $>$17.8 AB mag. & black  triangle    \\
\ W4(22\mum)$^1$ & $>$15.4 AB mag. & black  triangle    \\
\ 5.5 GHz$^2$ & F$_{\nu}$=46$\pm$16\muu Jy & black asterisk \\
\ 9.0 GHz$^2$ & F$_{\nu}$$<$18\muu Jy & black triangle \\
\ 5.5 GHz$^2$ & F$_{\nu}$$<$16\muu Jy & black triangle \\
\ 9.0 GHz$^2$ & F$_{\nu}$$<$18\muu Jy & black triangle \\
\
\ Outside the XRT &  &           \\
\ Galaxy D     &    &            \\
\ g'    $^2$  & 22.46$\pm$0.07 AB mag. & red circle \\
\ r'    $^2$  & 20.72$\pm$0.02 AB mag. & red circle \\
\ i'    $^2$  & 20.17$\pm$0.02 AB mag. & red circle \\
\ z'    $^2$  & 19.84$\pm$0.02 AB mag. & red circle \\
\ J     $^2$ & 18.69$\pm$0.02 AB mag. & red circle \\
\ H     $^2$ & 18.35$\pm$0.03 AB mag. & red circle \\
\ K$_s$ $^2$  & 18.06$\pm$0.05 AB mag. & red circle \\
\ 5.5 GHz$^2$ & F$_{\nu}$=151$\pm$23\muu Jy & red square \\
\ 9.0 GHz$^2$ & F$_{\nu}$=83$\pm$28\muu Jy & red square \\
\ W1(3.4\mum)$^1$ & 18.65$\pm$0.08 AB mag. &  red   star \\
\ W2(4.6\mum)$^1$ & 18.87$\pm$0.18 AB mag. &  red   star \\
\ W3(12\mum)$^1$ &$>$17.8  AB mag. &  red  triangle  \\
\ W4(22\mum)$^1$ &$>$15.4  AB mag. &  red  triangle  \\ \hline
\end{tabular}

$^1$ WISE satellite;
$^2$  ATCA

\end{table}

\begin{figure*}
\centering
\includegraphics[width=8.8cm]{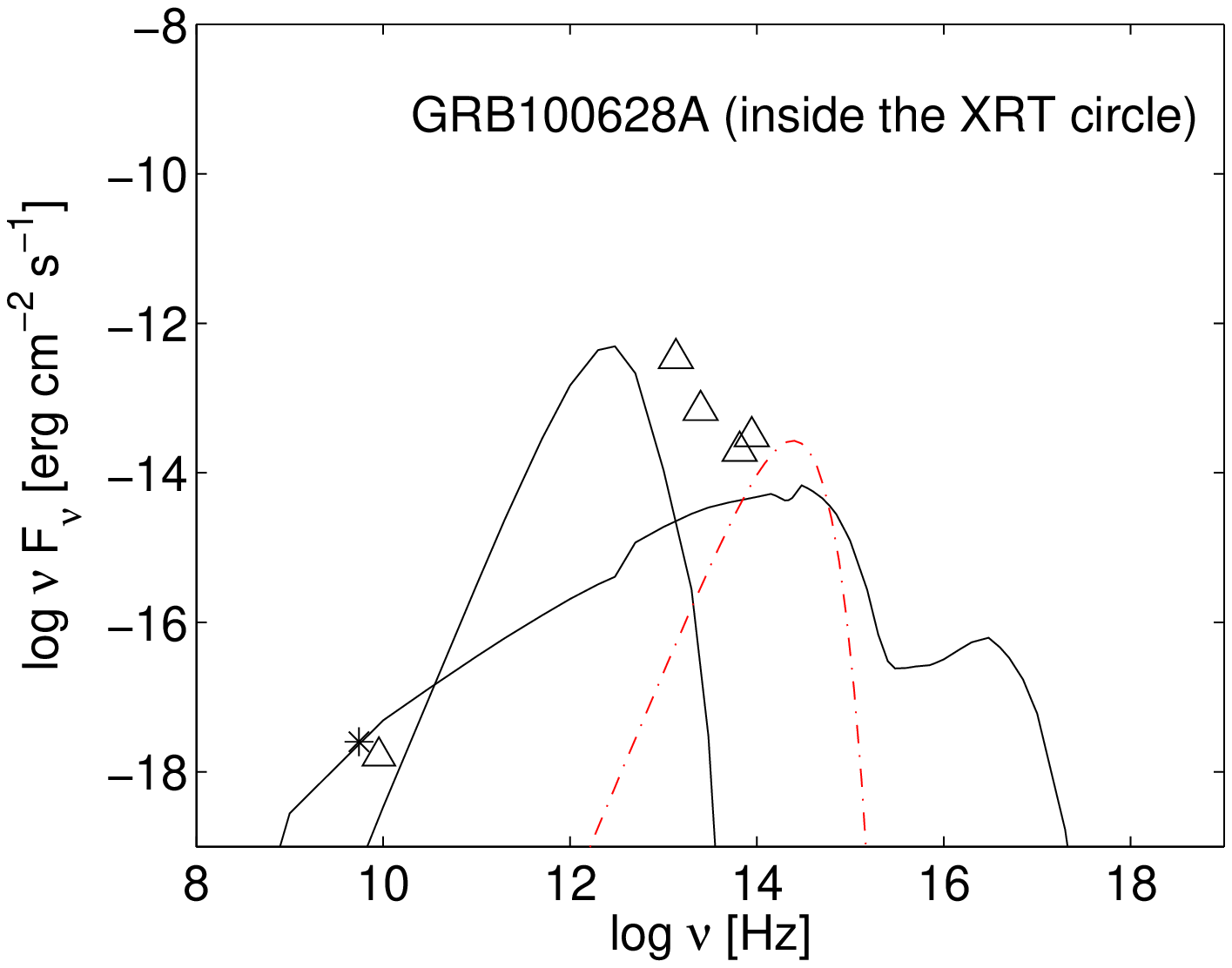}
\includegraphics[width=8.8cm]{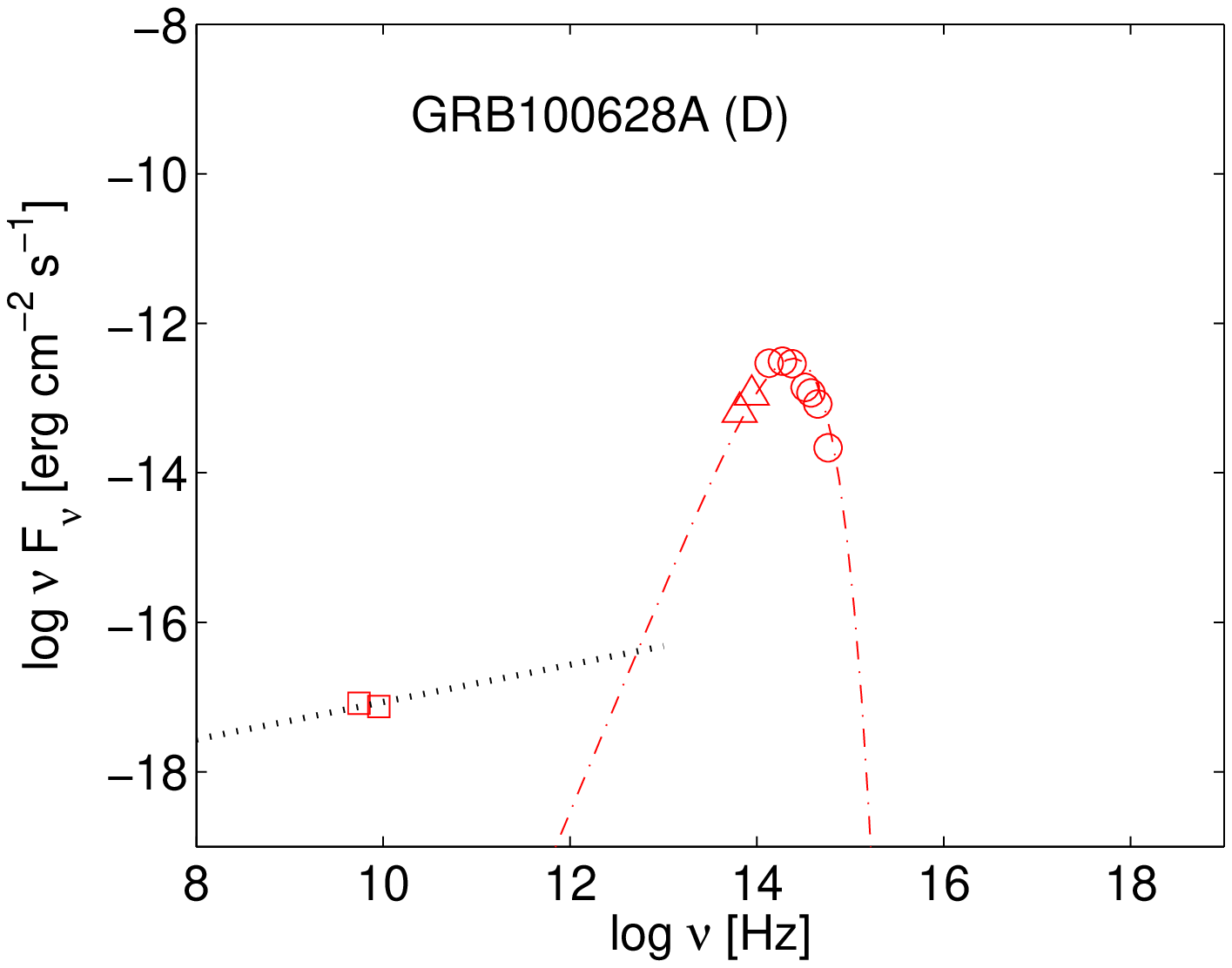}
\includegraphics[width=8.8cm]{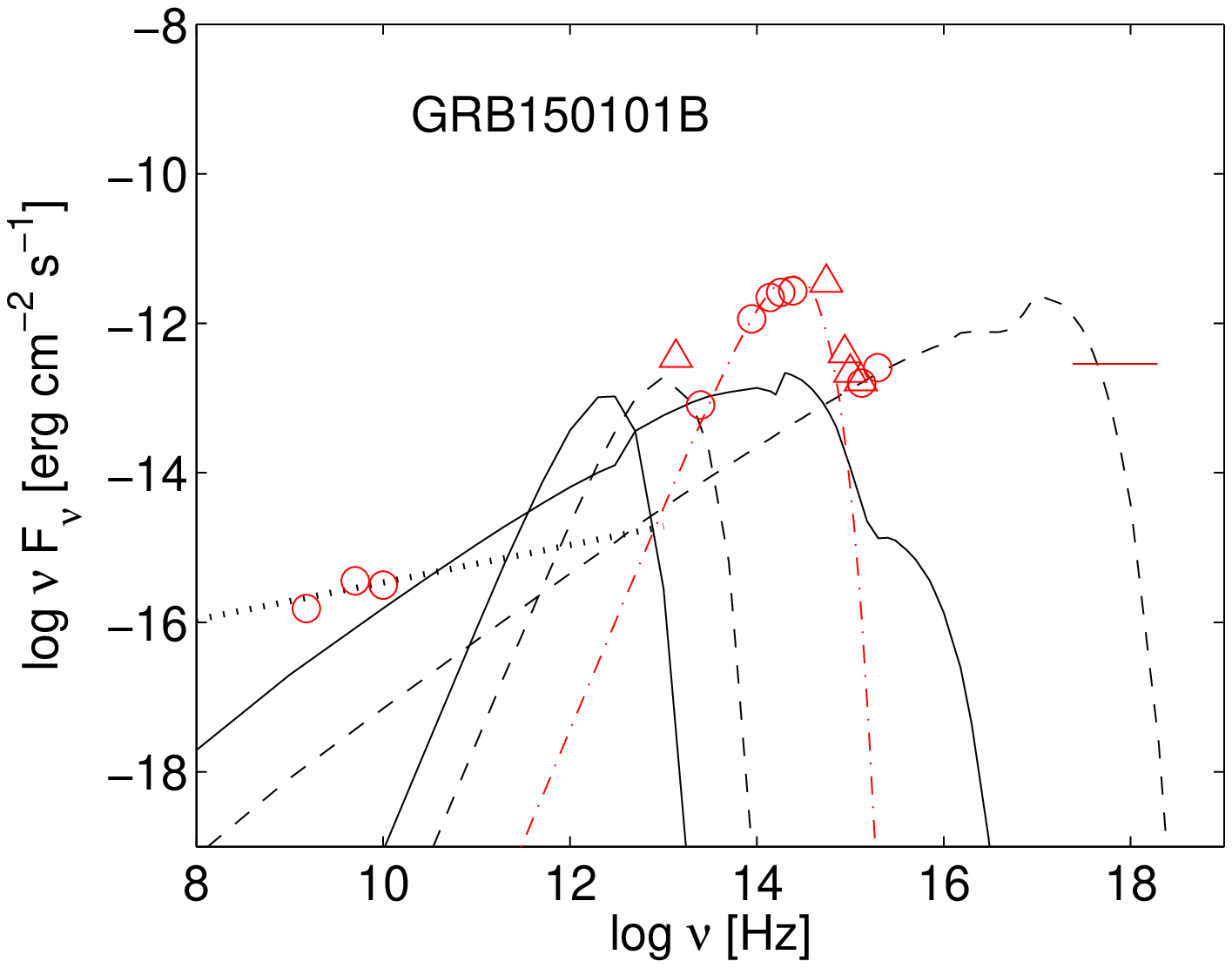}
\includegraphics[width=8.8cm]{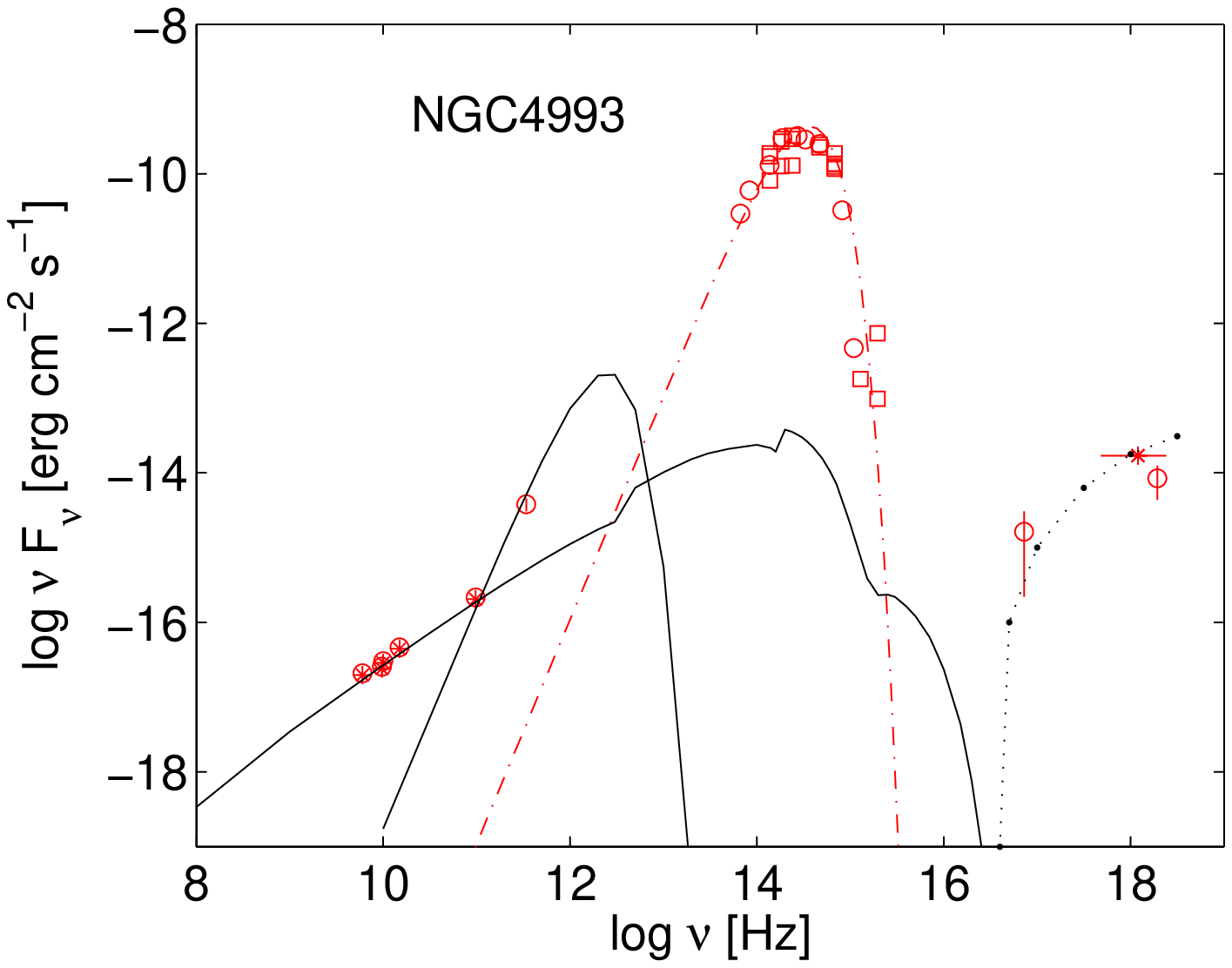}
 \caption{Top:  fit of the  SGRB100628A SED (data are described in Table 4).
Left. Galaxy C: solid black lines: mod$_{98}$; red dot-dashed line: underlying star flux;  
open triangles: upper limits; asterisk: bright source inside the XRT circle;
Right. Galaxy D: red circles (data) and red triangles (upper limits) from NG15;
red dash-dotted: underlying star flux; dotted line: radio synchrotron.
Bottom left: fit of SGRB150101B SED.
 Black dashed: shock dominated model with \Vs=700 \kms;
 black solid: mod$_2^*$;
red open circles: data from Xie et al; red open triangles:  upper limits; black dotted line:
synchrotron radiation; red horizontal segment: observed X-ray band.
Bottom right: fit of NGC4993 adapted from Contini (2018a).
}
\end{figure*}

 In Fig. 4  we present the modelling of   SGRB100628A continuum SED  outside and inside the XRT circle  
and we compare  them with the SED of SGRB150101B host  and NGC4993 galaxies.
Each model is represented in the diagrams  by two lines, one representing the bremsstrahlung
emitted from the nebula and the other the dust reprocessed radiation  consistently calculated (Sect. 2.3).
The lines are represented by the same symbol.  More clouds may   contribute to the observed SED,
therefore  different models can appear in  each galaxy diagram. In all the diagrams the black body radiation
from the underlying old stellar population  dominates the optical-IR range.

\subsubsection{Radio SED inside the XRT error circle}

A single relatively bright radio source inside the XRT error circle   was detected
slightly west of object  A (NG15 their fig. 2)  with an integrated density flux at 5.5 GHz of $F_{\nu}$=46$\pm$
16 $\mu$Jy. This source lies 2" north-east of blob 98, which  shows an  emission-line spectrum.
The radio emission is not detected at 9.0 GHz ($F_{\nu}$ $<$ 18 $\mu$Jy).
NG15 did not detect galaxy C as shown in  their fig. 3 in the radio, only upper limits at 5.5 GHz and 9.0 GHz.
With regards to the possible identity of the radio source inside the XRT circle, we agree with
NG15 who claim that the  radio emission is situated in the star forming
region of galaxy C tidal tail.
We  consider  blob 98 presented by NG15 in their fig. 3 (that is located  in the tidal tail of galaxy C)
because it was found that  $U$  is particularly high suggesting that blob 98 could be close to 
the SB bulk.
 The clouds throughout the host galaxy blobs emit  line  as well as  continuum radiation fluxes. 
The continuum  is   represented by the bremsstrahlung (Sect. 2.3)
which appears throughout the SED from radio to the soft X-ray frequencies, 
reradiation by dust in the IR, and the underlying old star radiation which dominates
in the IR-optical range.
Therefore, the continuum emitted from blob 98 - calculated by 
the same model (mod$_{98}$) that was found   by fitting the line ratios (Table 2)
appears in Fig. 4 (top-left) diagram.
In the radio range, the datum observed  at 5.5 GHz and the upper limit at 9.0 GHz are used to constrain
the continuum calculated  within the XRT circle. 
We explain the trend of the observed radio flux by  noticing that the two datapoints 
(one observed and one upper limit) belong to two different sources. The upper limit refers to dust reradiation 
in  the  far IR domain, while the observed datum at  a lower frequency  is fitted by  the thermal bremsstrahlung 
emitted from the gas.
 The data in the IR observed by the WISE satellite are all upper limits. They  suggest that the underlying
stellar population flux calculated by a temperature of 3000K is low but it is still   emerging over the
bremsstrahlung.

\subsubsection{Galaxy D. Underlying old star population flux outside the XRT}

 In  previous works (e.g. Contini 2018b) we have shown that the continuum SED is composed
by different contributions: the bremsstrahlung from the gas emitting the lines, reradiation from dust,
the underlying stellar background, radio synchrotron radiation, and eventually X-ray from the AGN. 
We suggest that galaxy D which does
not contribute to the observed optical lines, shows   the underlying old stellar population flux
observed throughout the continuum SED in the IR.
In Fig. 4 (top right diagram),  we  show  the  optical-IR continuum  observed from 
galaxy  D at z$\sim$ 0.3 (NG15, their fig. 7) outside the XRT error circle.
The optical $g'i'r'z'$ and NIR JHK bands  observed from galaxy D are well fitted  by
a black body flux at  $\sim$ 3000 K,
 confirming  that  in the elliptical galaxy D the stellar population is old.     
The galaxy is also detected at other frequencies.
The radio data observed from galaxy D   follow the  trend of the synchrotron radiation created by
the Fermi mechanism at the shock front.

\subsubsection{SEDs of SGRB150101B and NGC4993 hosts}

We  compare the continuum SED of GRB100628A  (inside and outside the XRT error circle)  
with the SEDs presented for other SGRB hosts enclosing an AGN.
Levan et al. (2015), Fong et al. (2015) and Troja et al. (2015)  suggested that
GRB150101B contains a central AGN. It was confirmed by Xie et al (2016)
multiwavelength analysis on the basis of  Chandra observations.
Xie et al adopted z=0.1345 as the redshift of the host galaxy. 
There are no line fluxes reported for SGRB150101B host galaxy. Modelling directly the continuum SED
in Fig. 4 (bottom left diagram)  we reproduce the observed X-ray band and the optical-UV data 
by the bremsstrahlung  calculated   adopting a
shock dominated  model ($F$=0, $U$=0) with \Vs=700 \kms and \n0=300 \cm3. 
This model also constrains the dust reradiation peak.
High velocities are confirmed by  collisionally heated grains to relatively high temperatures. 
In fact, the dust reradiation  peak is shifted towards relatively high frequencies.  
Radio emission is reproduced by  synchrotron radiation created by the Fermi mechanism at the shock front.
Besides the shock dominated model, we check whether the bremsstrahlung calculated by an AGN dominated model 
(+shock) may contribute to the continuum SED. 
An  AGN  model has been   selected among those fitting  SGRB100628A host spectra  (mod$_2^*$, Table 2).
Its contribution to the SED of SGRB150101B is  shown  at ${\nu}$$\sim$ 2.5$\times$10$^{13}$ Hz in the IR
and to the flux at $\nu$$\sim$ 10 GHz in the radio range.
The underlying stellar radiation  flux in the IR is reproduced by a black-body at 3000K.

With regards to NGC4993 which is reported in Fig. 4  bottom right  diagram,  
radio emission is thermal bremsstrahlung, while both the soft- and the hard- X-ray fluxes come from the AGN.
Comparison with the two SGRB host  (100628A and 150101B) SEDs confirms that   
radiation  from the underlying stellar population  - which is reproduced by a black-body at
4500K - is outstanding  because exceeding the underlying fluxes in the other hosts by
a factor of $\sim$100. (Contini 2018a).

\section{N/O abundance ratios}

\begin{table*}
\centering
\caption{AGN dominated models for Perley et al (2012) short GRB100206A host spectra at z=0.4068 (\Hb=1)}
\begin{tabular}{lcccccccccccccccc} \hline \hline
\                 & [OII] &  [OIII]  & \Ha  & [NII]   & [SII]  &[SII] &\Vs & \n0 & $D$ & N/H & O/H & S/H &$F$\\
\                 &3727   &   5007+  &6563  & 6548+   & 6716   & 6731 & 1  & 2   & 3   & 4   & 5   & 6   & 7\\ \hline
\ subtr           &1.34   &   0.23   &3.    &1.58     & 0.53   &0.35  & -  & -   & -   & -   & -   & -   &   \\
\ modp1*          &1.8    &  0.26    &3.    &1.4      & 0.53   &0.4   &100 &100  &60   &7.9  &8.78 &6.3  &3  \\
\ North           & 0.9   & 0.127    & 3    & 1.52    & 0.39   & 0.36 & -  & -   & -   & -   & -   & -   & - \\
\ modp2*          & 1.1   &0.13      &3.47  & 1.52    &0.39    & 0.53 &200 &240  &6    &8    &8.78 &7.9  &1.6 \\
\ Center          & 1.57  &0.27      &3.    & 1.67    & 0.15   &0.149 & -  & -   & -   & -   & -   & -   & -  \\
\ modp3*          & 1.7   &0.27      & 3    &1.69     &0.12    &0.17  &200 &240  &2.6  &7.92 &8.78 &6.3  &2  \\
\ South           & 1.6   &0.36      &3.    &1.8      &0.5     &0.81  &-   & -   & -   & -   & -   & -   & -  \\
\ modp4*          & 1.6   &0.36      & 3.5  &1.82     &0.54    &0.75  &200 &240  &2.   & 8   &8.78 & 7   &2.4  \\ \hline
\end{tabular}

1:\kms; 2:\cm3; 3:10$^{18}$cm; 4:12+log(N/H); 5:12+log(O/H); 6:12+log(S/H);
7: 10$^8$ photon cm$^{-2}$ s$^{-1}$ eV$^{-1}$ at the Lyman limit

\centering
\caption{AGN dominated models for SGRB130603B  at z=0.356 and for SGRB051221a  z=0.546 host spectra (\Hb=1)}
\begin{tabular}{lcccccccccccccccc} \hline \hline
\            & [OII] & \Hg & [OIII] & \Ha &[NII] &[SII]&[SII] &\Vs &\n0 & $D$ & N/H & O/H & S/H & $F$ \\
\            &3727+  &     & 5007+  &     &6585  &6717 &6731  & 5  & 6  &  7  &  8  &  9  & 10  & 11   \\ \hline
\ OT site$^1$& 4.47  &-    & 0.87   & 3   &0.7   &0.66 &0.33  & -  & -  &  -  &  -  &  -  &  -  &  -   \\
\ mS1*       & 4.47  &0.46 & 0.84   & 2.98&0.8   &0.56 &0.46  &115 &220 &5    &7.3  &8.83 &6.15 &    11\\
\ OT site$^2$& 3.4   &0.89 &0.76    & 3   &0.57  &1.19 &0.59  &  - & -  &  -  &  -  & -   &  -  &   -  \\
\ mS2*       & 3.4   &0.457&0.74    &3    &0.58  &0.85 &0.7   &120 &220 &5.6  &7.47  &8.83&6.4  &   8\\
\ core$^2$   &3.55   &0.49 &0.75    & 2.96&0.85  &0.63 &0.5   & -  &  - &  -  &  -  &  -  &  -  &  -  \\
\ mS3*       &3.4    &0.46 &0.74    & 3   &1.2   &0.85 &0.7   &120 &220 &5.6  &7.6  &8.83 &6.4  &8  \\
\ arm$^2$    &4.3    &[0.5]&0.85    & 3   &1.8   &0.27 &0.5   &  - &  - &  -  &  -  &  -  &  -  &  -  \\
\ mS4*       &4.47   &0.46 &0.84    & 2.98&0.8   &0.56 &0.46  &120 &220 &5.0  &7.3  &8.83 &6.15 &11\\
\ obs$^3$    &3.05   &-    &0.77    &3    &0.78  & -   & -    &  - &  - &  -  &  -  &  -  &  -  &  -  \\
\ mS5*       &3.4    &0.46 &0.74    &3    &1.2   &0.85 &0.7   &120 &220 &5.6  &7.6  &8.83 &6.4  &8  \\
\ 051221a$^4$&7.8    &0.46 &5.17    &3    & -    & -   &  -   &  - &-   &-    &-    &-    &-    &-    \\
\ mS6*       &7.9    &0.46 &5.1     &2.96 &0.54  &0.27 &0.28  &170 &100 &1    &7.   &8.81 &6.15 & 24.5\\ \hline
\end{tabular}

$^1$de Ugarte Postigo et al. (2014)(X-shooter); $^2$de Ugarte Postigo et al. (2014)(FORS);$^3$ Cucchiara et al. (2013)(DEIMOS);
$^4$ Soderberg et al. (2006)(Gemini-N GMOS);
5:\kms; 6:\cm3; 7:10$^{18}$cm; 8:12+log(N/H); 9:12+log(O/H); 10:12+log(S/H);
11:10$^8$ photon cm$^{-2}$ s$^{-1}$ eV$^{-1}$ at the Lyman limit

\centering
\caption{AGN dominated models for Berger  (2009) sample of short GRB host spectra (\Hb=1)}
\begin{tabular}{lcccccccccccccccc} \hline \hline
\  GRB    &z      & [OII] &  [OIII]  & \Ha  & [NII]   &\Vs & \n0 & $D$ & N/H & O/H  &$F$\\
\         &       &3727   &   5007+  &6563  & 6548+   & 1  & 2   & 3   & 4   & 5    & 6\\ \hline
\ 061006  &0.4377 &4.24   &   1.89   &-     &-        & -  & -   & -   & -   & -    & - \\
\ modb1*  &       &4.5    &  1.8     &3.    &-        &100 &100  &0.8  &-    &8.82  &0.2\\
\ 061210  &0.4095 & 1.71  & 1.6      & 3    & 0.15    & -  & -   & -   & -   & -    & - \\
\ modb2*  &       & 1.8   & 1.6      &3.    & 0.15    &180 &350  &0.16 &6.9  &8.8   &0.06\\
\ 070724  &0.4571 & 2.47  &1.13      &3.    & -       & -  & -   & -   & -   & -    & -  \\
\ modb3*  &       & 2.20  &1.14      & 3    & -       &130 &350  &0.34 & -   &8.81  &0.8\\
\ 050709  &0.1606 &  -    &3.7       &3.    &0.29     &-   & -   & -   & -   & -    & -  \\
\ modb4*  &       &(1.9)  &3.6       & 3.   &0.32     &80  &200  &80   &7.25 &8.6   &40   \\ \hline
\end{tabular}

1:\kms; 2:\cm3; 3:10$^{18}$cm; 4:12+log(N/H); 5:12+log(O/H);
6: 10$^8$ photon cm$^{-2}$ s$^{-1}$ eV$^{-1}$ at the Lyman limit

\end{table*}


Fig. 5  shows the distribution of N/O relative abundances calculated for different types of galaxies as a function of the
redshift. The results obtained for the SGRB100628A blobs are compared with those of other SGRB
host galaxies, LGRB hosts, AGN, LINER,  and HII regions at z$\leq$0.4.
In a previous paper  (Contini 2018a) we have calculated N/O for NGC4993
by an AGN dominated model which was constrained  by  reproducing the observed continuum SED.
For the other SGRB host  galaxies we have used SB models because well fitting the
line ratios. Moreover, a SB regime is  generally recommended in GRB hosts.
Fig. 5 shows that although the N/O  ratios calculated for SGRB100628A are in agreement with those calculated
for AGNs and other objects at the same z,
N/O ratios  calculated   from the Perley et al. (2012) spectra  of the
SGRB100206A host complex at z=0.4068  (Fig. 5, large squares)  exceed the solar values in all the observed positions (Contini 2018a,  fig. 6).
We  would like to investigate whether  the modelling of the  line  ratios  adopting a power-law  flux
could lead to different results, in particular for N/O.
The  AGN hypothesis  was  discarded by Perley et al. on the basis of the  \Ha and [OIII] line strengths,
however, by qualitative analysis.
 We checked the AGN model  also for the SGRB100206A host complex (Perley et al (2012), the Berger (2009) host 
sample at z=0.16-0.45, for the host galaxies of
SGRB130603B (de Ugarte Postigo et al. 2014) at z=0.356 and for the SGRB051221a host (Soderberg et al. 2006) at z=0.546.
The corrected observed line ratios  are reproduced by AGN dominated models (modp1*-modp4*,
mS1*-mS4*, modb1*-modb4*)  (Tables 4-6).
The AGN models are described in  the last seven columns of the tables.
Tables 4 - 6 show that the O/H values calculated by the AGN models are  close to the  solar ones and the
N/H are nearly solar for SGRB100206A, $\sim$0.3 solar for SGRB130603B and SGRB051221a and $\sim$0.1 solar for Berger (2008) sample.
Therefore, they are dislocated throughout the BPT diagram (Fig. 3).
Only SGRB100206A and SGRB061210 seem  to be SB dominated.
The \Hb fluxes calculated at the nebula by AGN models are compared  in Table 8 with
 \Hb fluxes calculated by SB models by Contini (2018a).
Table 8  shows that the  \Hb absolute fluxes calculated by an AGN model
exceed  those calculated by a SB  for Perley et al.,
de Ugarte Postigo et al. and   Soderberg et al. objects, but are lower for the Berger sample, except for SGRB050709
which, however, is less reliable because the spectra lack the observed [OII]/\Hb line ratio.
We conclude that both an AGN and a SB can  photoionize and heat the SGRB host gas.

\begin{figure}
\centering
\includegraphics[width=9.6cm]{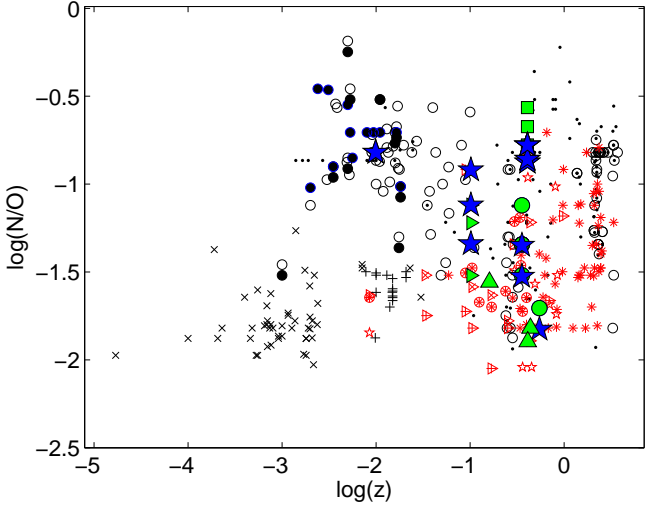}
 \caption{Distribution of N/O throughout the redshift.
 SB models for SGRB hosts- represented by  filled green symbols-  were calculated by Contini (2018a); 
large squares: SGRB100206A host at different locations (Perley et al 2012); 
large triangles: SGRB host sample at different z (Berger 2009);
large  circles: SGRB130603B host at different sites (de Ugarte Postigo at al 2014) 
and SGRB051211a host (Soderberg et al 2006);
small triangles: SGRB100628A blobs calculated by SB models (present paper, Tables 1 and 2).
Large filled blue stars: AGN model results for NGC4993 (Contini 2018a),  
SGRB100628A blobs (present paper, Tables 1 and 2), SGRB100206A, SGRB130603B and SGRB051221a hosts (present paper, 
Tables 4-6).
Other symbols are described in  Table 7.
}
\end{figure}

\begin{table}
\centering
\caption{Symbols in Fig. 5}
\begin{tabular}{llcl} \hline  \hline
\ symbol   &object                       & Ref. \\ \hline
\ red  asterisks &GRB hosts                   & (1) \\
\ red  triangle+cross    & LGRB hosts         & (2)  \\
\ red  pentagrams  & LGRB different hosts     &  (3)  \\
\ red triangle +plus& LGRB hosts with WR stars& (4) \\
\ red encircled asterisks &LGRB at low z     & (5)\\
\ black  dots & starburst galaxies              & (6,7)\\
\ black open circles & AGN                          & (7,8,9)  \\
\ black filled circles & LINER                 & (10) \\
\ black plus &  low-luminosity nearby galaxies & (11)  \\
\ black cross& HII regions in local galaxies   & (12)  \\ \hline
\end{tabular}

(1) (Kr\"{u}hler et al.);
(2) (Savaglio et al.);
(3) (Contini 2016a, table 8);
(4) (Han et al.);
(5) (Niino et al.);
(6), (7) (reported by Contini 2014 and references therein);
(8) (Contini 2016a);
(9) (Koski 1978, Cohen 1983, Kraemer et al. 1994, Dopita et al. 2015);
(10) (Contini 1997);
(11) (Marino et al. 2013);
(12) (Berg   et al. 2012);
\end{table}

In Fig. 5, N/O results  of AGN models  for the SGRB100206A spectra,  
 for   SGRB130603B  and for  SGRB051221a
are   superimposed on  previous results obtained 
by SB models.
The N/O ratios calculated by the  AGN models roughly follow and complete the AGN trend.
The lowest  N/O ratios  calculated by  SB   models for the Berger (2009) SGRB host sample appear
in  the N/O domain  of LGRB hosts  at z$\leq$0.4.

\section{Concluding remarks}

NG15  presented observations of line and continuum fluxes  from the   galaxy complex
inside and outside the XRT error circle of SGRB100628A,  in order to   reveal the 
host galaxy.  They indicated after a detailed discussion  galaxies C and D as the most appropriate ones.
They  add that the "higher redshift
solution (for galaxy D) fits slightly better into the ensemble properties of short GRBs known so far."
This  claim  will perhaps  change  by future observations.

In particular, the [OII] lines were provided by the observations only for blobs 1 and 2. For the other blobs
the [OII] lines were excluded from the spectra by the use of the GG 435 filter which cuts off
 the lines at $\lambda$$<$ 4400\AA.
 The [OII] lines together with the [OIII]5007+ lines are basic to constrain the models.
 We consider two extreme cases. Assuming that the [OII] line fluxes   are of the same order  as those 
 observed in blobs 1 and 2
we can predict, qualitatively,  that the physical conditions  throughout galaxy C are nearly homogeneous.
As a second choice  we assume  that the  [OII] line  fluxes  are  very low, therefore  we  can proceed by
the detailed modelling of the spectra.
We consider in the first instance that  the  stars are the dominant illumination source, therefore  
we have used in our model a black-body
photoionizing radiation. In a turbulent regime   such as that acting within the GRB dominated hosts
radiation is coupled to  the  collisional effects due to shocks which most likely derive  from the progenitor
merging.
We have found that for blobs 1 and 2
which show low [OIII]/[OII] line ratios, $D$   and $U$ are  much lower than for blobs 3, 4, 113, 117, 
which are aligned with the NE/SW slit. 
Blob 98  which is located in the tidal tail of the galaxy has the  maximum \Ts=10$^5$K and $U$=10. 
However, the inhomogeneous  distribution of   $U$ throughout galaxy C is not explained by the SB dominated models.
This  suggests that   a different  photoionizing  source should contribute
to the calculations of the line intensities. 
Therefore, we have modelled the line ratios observed from the blobs  by an  AGN dominated model.
The AGN  flux  which yields the best fit is similar to that found in the NLR of  Seyfert 2 galaxies (log$F$$\sim$11-12).
Shock velocities and pre-shock densities are suitable to those found in  the AGN NLR.
An  AGN and a SB can coexist in galaxy C as was suggested e.g. by Contini (2016b) for galaxies at z$\sim$0.8.
The merger-AGN connection in galaxies at z=0 was  recently discussed  by Ellison et al (2019).
On the basis of the calculated \Hb line flux at the nebula, 
 the AGN dominates the spectra from blobs 
3, 4 , 113, 117, which are displaced throughout galaxy C  along the axis of the bicone AGN structure 
in the  North-East/South-West slit direction (Fig. 1),
while the SB bulk is close to blob 98 in the tidal tail of the galaxy, in agreement with NG15.
Blob 98 is also reached by the AGN flux.
Blobs 1 and 2 are located in the ISM outside the AGN photoionization cones. They roughly indicate that
 the apex of the AGN  bicone should be   near blobs 117 and 3.
 Fragmentation is maximum near blobs 1 and 2.
O/H are nearly  solar  for all the gaseous clouds in  the blobs, while N/H ratios span  a factor  of $\leq$10. 

Radio observations reported
 a point source  at ${\nu}$=5.5GHz and an upper limit
at 9 GHz inside the XRT error circle,  close to blob 98. They are explained by the SB model calculated for blob 98.
In fact we  calculated    the continuum SED by the  same model  which reproduces the line ratios
observed  from the starburst in  galaxy C tidal tail.
The results show  that the two frequency emissions have  different origins.
One  is bremsstrahlung from  the  gas, whereas the  other (the upper limit) is explained
 by dust reprocessed radiation  which decreases rapidly at low frequencies in the far IR. 
 Galaxy D outside the XRT error circle  shows
 the  flux from the underlying old star  population at T=3000 K.
 The radio data in galaxy D follow the synchrotron radiation trend.
We compare the SGRB100628A host SED with  that of SGRB150101B  host and  of NGC4993 which encloses an AGN. 
The continuum SED of SGRB150101B host is reproduced  by a  shock dominated model
 calculated by \Vs=700 \kms and \n0=300 \cm3 (Fig. 4) which explains the X-ray band, the optical-UV  data,
relatively hot dust reradiation and  radio synchrotron radiation created by the Fermi mechanism.
Contini (2018a)  confirmed that  the line ratios  observed from NGC4993 
the host galaxy of GW170817 can be  explained by  an AGN (Fig. 4). 
Comparison with the SGRB100628A and SGRB150101B SEDs   strengthens the evidence 
  that the underlying stellar radiation   throughout the NGC4993  host  SED is outstanding
  because it exceeds the IR flux in other objects  by at least  a factor of 100.

New  calculations  by AGN dominated models of a  sample of SGRB host galaxy spectra, 
that were previously  modelled adopting  a SB photoionizing source, are presented.
The  line ratios calculated by the AGN source nicely  fit the observed ones,
as  well as those  calculated by  SB dominated models (Contini 2018a). 
 The precision of model calculations is cross-checked with the observations.
More spectroscopic data would better  constrain the models. 
 The BPT diagram (Fig. 3) shows ambiguous results because the N/H ratios
span a large range of values  (Kewley et al 2001).
Therefore, we  have selected SB from AGN dominated hosts comparing  the calculated \Hb line flux intensity (Table 8). 
Summarizing,  model results suggest that an AGN  and a SB  may coexist in  SGRB100628A.  
Updating  the diagram  that shows  N/O versus the redshift  in the light of the new results (Fig. 5), we obtain that
N/O relative abundances  in SGRB host galaxies  follow the AGN distribution, while
 NGC4993  at z=0.009873  is seen at the  edge of the AGN-LINER domain  of local galaxies.
A SB dominates  the hosts of the Berger  (2009) SGRB  galaxy sample.
 A more  significant  N/O trend distribution as a function of z on a large scale
will  appear when  spectral data for SGRB hosts at z$>$1
will be available.

\begin{table}
\centering
\caption{\Hb fluxes calculated  by AGN and SB dominated models}
\begin{tabular}{lcccccccccccccccc} \hline \hline
\  SGRB              & \Hb(AGN) & \Hb(SB) \\ \hline
\ 130603B OT  site1 $^1$         & 0.0243   & 0.014   \\
\ 130603B OT site2  $^1$        & 0.022    & 0.016   \\
\ 051211a $^2$          & 0.01     & 0.007   \\
\ 100206A subtr $^3$            & 0.052    &0.003    \\
\ 100206A North  $^3$           & 0.054    &0.025    \\
\ 100206A Center $^3$           &0.024     &0.01     \\
\ 100206A South  $^3$           &0.018     &0.009    \\
\ 061006 $^4$           &0.0015     &0.006   \\
\ 061210 $^4$           &0.006     &0.167    \\
\ 070724 $^4$           &0.0065    &0.072    \\
\ 050709 $^4$           &0.42      &0.19     \\ \hline
\end{tabular}

 $^1$ de Ugarte Postigo et al. (2014); $^2$ Soderberg et al. (2006);
$^3$ Perley et al. (2012); $^4$ Berger (2009).

\end{table}

\section*{Acknowledgements}

 I  am grateful to Prof Forveille, the Editor-in-Chief of A\&A for permitting
to reproduce  Nicuesa Guelbenzu et al. (2015, fig. 2).

\section*{References}

\def\ref{\par\noindent\hangindent 18pt}

\ref Baldwin, J., Phillips, M., Terlevich, R. 1981, PASP, 93, 5
\ref Barthelmy, S.D. et al 2005, Space Sci. Rev., 120, 143
\ref Berg, D.A. et al 2012, ApJ, 754, 98
\ref Berger, E. 2014, ARA\&A, 52, 43
\ref Berger, E. 2013, ApJ, 765, 121 
\ref Berger, E. 2010, ApJ, 722, 1946
\ref Berger, E. 2009, ApJ, 690, 231
\ref Berger, E. et al. 2005, Nature, 438, 988
\ref Bloom, J.S. et al. 2006 ApJ,638, 354 
\ref Cohen, R.D. 1983, ApJ, 273, 489
\ref Contini, M. 2018b, arXiv180103312
\ref Contini, M. 2018a, A\&A, 620, 37
\ref Contini, M. 2016a, MNRAS, 460, 3232
\ref Contini, M. 2016b, MNRAS, 461, 2374
\ref Contini, M. 2014, A\&A, 564, 19
\ref Contini, M. 2012, MNRAS, 425, 1205
\ref Contini, M. 1997, A\&A, 323, 71
\ref Contini, M. \& Viegas, S.M. 2001a ApJS, 137, 75
\ref Contini, M. \& Viegas, S.M. 2001b ApJS, 132, 211
\ref Contini, M., Radovich, M., Rafanelli, P., Richter, G.M. 2002, ApJ, 572, 124
\ref Cox, D..P. 1972, ApJ, 178, 159
\ref Cucchiara, A. et al 2013, ApJ, 777, 94
\ref de Ugarte Postigo, A. et al. 2014, A\&A, 563, 62 
\ref Dopita, M.M. et al. 2015, ApJS, 217, 12
\ref Draine, B.T. \& Lee, M.M. 1994, ApJ, 285, 89
\ref Draine, B.T. \& Salpeter, E.E.  1979, ApJ, 231, 438
\ref Dwek, E. \& Cherchneff, I. 2011 ApJ, 727, 63
\ref Ellison, S.L. et al ArXiv:1905.08830
\ref Ferland, G. 2016 arXiv160308902,	
	The Lexington Benchmarks for Numerical Simulations of Nebulae
\ref Fong, W. et al. 2013, ApJ 769, 56
\ref Fong, W. , Berger, E. 2013, ApJ, 776, 18
\ref Fong, W., Berger, E., Fox, D.B. 2010, 708, 9
\ref Gehrels, N., White, N., Barthelmy, S. et al. 2005, Nature, 437, 851
\ref Grevesse, N., Sauval, A.J. 1998 SSRv, 85, 161
\ref Han, X. H., Hammer, F., Liang, Y. C., Flores, H., Rodrigues, M., Hou, J. L., Wei, J. Y.
 2010, A\&A, 514, 24
\ref Helou, G. 1986, ApJ, 311, L33
\ref Ho, L., Filippenko, A. V., Sargent, W.L.W. 1993, ApJ, 417, 63
\ref Kewley, L.J., Dopita, M.A., Sutherland, R.S., Heisler, C.A., Trevena, J. 2001, ApJ, 556, 121
\ref Koski, A. 1978, ApJ, 223, 56
\ref Kouvelioutou, C et al. 1993, ApJL, 413, L101
\ref Kraemer, S.B., Wu, C.-C., Crenshaw, D.M.,Harrington, J.P. 1994, ApJ, 435, 171
\ref Kr\"{u}hler, T. et al. 2015 A\&A, 581, 125
\ref Immler, S., Starling, R.L.C., Evans, P.A., Barthelmy, S.D., Sakamoto, T. 2010, GCN Rep. 290, 1 
\ref Levan, A.J. et al. 2017 ApJL, 848, L31 
\ref Marino, R.A. et al. 2013, A\&A, 559, 114
\ref Mulchaey, J.S., Wilson, A.S., Tsvetanov, Z. 1996, ApJ, 467, 197
\ref Nicuesa Guelbenzu A. et al 2015, A\&A, 583, A88,  NG15
\ref Niino, Y. et al. 2016 Publ. Astron. Soc. Japan, 69, 27
\ref Osterbrock, D.E. 1974, in Astrophysics of Gaseous Nebulae, W.H Freeman and Co., San Francisco
\ref Pan, Y.-C. et al. 2017, ApJ, 848, L30
\ref Perley, D.A. et al. 2012, ApJ, 758, 122
\ref Rigby, J.R. \& Rieke, G.H. 2004, ApJ, 606, 237
\ref Savaglio, S., Glazerbrook, K., Le Borgne, D. 2009, ApJ, 691, 182
\ref Schmitt, H. R., Donley, J.L., Antonucci, R.R.J., Hutchings, J.B., Kinney, A.L., Pringle, J.E., 2003, ApJ, 597, 768
\ref Selsing, J. et al. 2018, A\&A, 616, 48
\ref Soderberg, A.M. et al. 2006, ApJ, 650, 261
\ref Starling, R.L.C., Beardmore, A.P., Immler, S. 2010, GRB coordinates Network, 10907,1
\ref Troja, E., Sakamoto, T., Lien, A., Cenko, S.B., Gehrels, N. 2015, GCN, 18289, Q   2015
\ref van den Bergh, S., Marshner, A.P., Terzian, Y  1973 ApJS, 26, 19 
\ref Villar, V.A. et al 2017, ApJ,851, L21
\ref Wu, Q., Feng, J., Fan, X. 2018, ApJ, 855, 46
\ref Xie, C.,  Fang, T.,  Wang, J., Liu, T., Jiang, X. 2016, ApJL, 824, L17

\end{document}